\documentstyle[aps,prl]{revtex}

\title{Bose-Einstein condensates with vortices in rotating traps}
\author{Y. Castin${}^1$ and R. Dum$^{1,2}$ \\
${}^1$Laboratoire Kastler Brossel$^*$, \'Ecole normale sup\'erieure,
24 rue Lhomond, F-75231 Paris Cedex 05, France \\
${}^2$ Institut d'optique, BP 147, F-91403 Orsay Cedex, France }
\date{16 March 1999}

\begin{document}
\maketitle

\newcommand{\sech}{\mbox{sech}}

\abstract{We investigate minimal energy  solutions 
with vortices for an interacting Bose-Einstein
condensate in a rotating trap. The atoms are strongly confined along
the axis of rotation $z$, leading to an effective 2D situation
in the $x-y$ plane. We first use
a simple numerical algorithm converging to local minima of
energy. Inspired by the
numerical results we present a variational
Ansatz in the regime where the interaction
energy per particle is stronger than the quantum of vibration in the harmonic
trap in the $x-y$ plane, 
the so-called Thomas-Fermi regime. This Ansatz allows an easy
calculation of the energy of the vortices as function of the
rotation frequency of the trap; it gives a physical understanding
of the stabilisation of vortices by rotation of the trap
and of the spatial arrangement of vortex cores. 
We also present analytical results
concerning the possibility of detecting vortices by a time-of-flight 
measurement or by interference effects. In the final
section we give numerical results for a 3D configuration.}

\section{Introduction}

After the achievement  of Bose-Einstein condensates in trapped atomic
gases \cite{BEC} many properties of these systems have been
studied experimentally and theoretically
\cite{REVUE}.  However a striking feature of superfluid helium,
quantized vortices \cite{ONSAGER},\cite{DONNELLY}, 
has not yet been observed in trapped atomic gases.
There is an abundant literature on vortices in helium II,
an overview is given in \cite{DONNELLY}.

The atomic gases have interesting properties which
justify efforts to generate vortices in these systems:
the core size of the vortices is adjustable, as
in contrast to helium the strength of the interaction can
be adjusted through the density;
the number of vortices in atomic gases
can be in principle well controlled;
for a small number of particles in the gas 
metastability of the vortices can be studied, that is one can watch
spontaneous transitions between configurations with different 
number of vortices.

Several ways to create vortices in atomic gases have been suggested. 
A method inspired from liquid helium consists in rotating the trap
confining the atoms
\cite{LEGGETT}; at a large enough rotation frequency  it becomes
energetically favorable at low temperatures to produce vortices; two different
paths could be in principle followed: (1) producing first a condensate 
then rotating the trap, or (2) cooling the gas directly in a rotating trap.
It has been recently proposed in \cite{BERRY} to use quantum topological
effects to obtain a vortex.
Other methods that do not rely on thermal
equilibrium have been suggested \cite{GENERATE1},\cite{GENERATE2}.

Here we study theoretically the minimal energy configurations 
of vortices in a rotating trap
\cite{HESS}. The model is defined in section \ref{sec:model}; 
in sections \ref{sec:prop} to \ref{sec:intuitive}
we assume a strong confinement
of the atoms along the rotation axis $z$ so that we face an
effective 2D problem in the transverse plane $x-y$.
We present numerical
results for solutions with vortices that are
local minima of the Gross-Pitaevskii energy functional 
(section \ref{sec:prop}). 
These solutions contain only vortices with a charge
$\pm 1$, the vortices with a charge larger than or equal to
2 are thermodynamically unstable (section \ref{sec:why}). 
We discuss possibilities
to get experimental evidence of vortices in atomic gases
in section \ref{sec:how}. Finally, we concentrate on the
regime where the interaction energy is much larger than the trap
frequencies $\omega_{x,y}$, the so-called Thomas-Fermi
limit \cite{REVUE}. This is complementary to the
work of \cite{ROKHSAR}. We obtain in this \lq\lq strong interacting" regime
analytical predictions based on a variational Ansatz that reproduce
satisfactorily the numerical results (section \ref{sec:intuitive}).
In section \ref{sec:3d}  we present results for vortices
in 3D, that is in a trap with a weak confinement along the
rotation axis.

\section{Model considered in this paper} \label{sec:model}

The atoms are trapped in a potential rotating at angular velocity
$\Omega$. In the laboratory frame the Hamiltonian of the gas
is therefore time dependent.
To eliminate this time dependence we introduce
a rotating frame at the angular velocity $\Omega$ so that the trapping
potential becomes time independent; this change of frame 
is achieved by the single-atom unitary transform:
\begin{equation}
{\cal U}(t)=e^{i\vec{\Omega}\cdot\vec{L} t/\hbar}
\end{equation}
where $\vec{L}$ is the angular momentum operator of a single atom.
As the unitary transform is time dependent the Hamiltonian in the
rotating frame contains an extra inertial term, given for each atom by
\begin{equation}
i\hbar {\cal U}^{\dagger}(t){d\over dt}{\cal U}(t) = -\vec{\Omega}\cdot\vec{L}.
\end{equation}

The atoms are interacting via the effective low energy potential
commonly considered in the literature,
$V(\vec{r_1}-\vec{r_2})=g_{3D}\delta(\vec{r_1}-\vec{r_2})$,
where the coupling constant is related to the $s$-wave scattering length $a$
(taken here to be positive)
and to the atomic mass $m$ by $g_{3D}=4\pi\hbar^2a/m$.
In this paper we consider the case 
of a dilute gas (with a density much smaller than $a^{-3}$) at 
zero temperature. We can
then assume that the $N$ particles of the gas are condensed in the same
state $\phi$. The wavefunction $\phi(\vec{r})$ is time
independent as we are in the rotating frame and minimizes the energy 
per particle in
the condensate, given by the Ginzburg-Landau
energy functional:
\begin{equation}
E[\phi,\phi^*] = \int d^3\vec{r}\ 
{\phi^*(\vec{r}\,)\left[ {\cal H}_0-\vec{\Omega}\cdot\vec{L} 
\right]\phi(\vec{r}\,)
\over \langle\phi|\phi\rangle} +{1\over 2}
{Ng_{3D} |\phi|^4 \over \langle\phi|\phi\rangle^2}.
\label{eq:ener3d}
\end{equation}
In this energy functional ${\cal H}_0$ contains the kinetic energy 
and the trapping potential energy of the particles:
\begin{equation}
{\cal H}_0=-{\hbar^2\over 2m}\Delta + U(\vec{r},t=0).
\end{equation}
The energy functional includes also the inertial term 
$-\vec{\Omega}\cdot\vec{L}$ and a term proportional to $|\phi|^4$ 
describing the interaction energy between the particles in the
mean field approximation.

As done in the present experiments with atomic gases we take 
the case of a harmonic trap, with eigenfrequencies $\omega_{\alpha}$:
\begin{equation}
U(\vec{r},t=0) = \sum_{\alpha=x,y,z}{1\over 2}m\omega_\alpha^2r_\alpha^2.
\end{equation}
Furthermore in all but in section \ref{sec:3d}
we will assume that the trapping potential is much stronger
along the $z$ axis than along the $x,y$ axis, 
with an oscillation frequency much larger 
than the typical interaction energy $Ng_{3D}|\phi|^2$ per particle.
This situation, although not realized experimentally yet, is not
out of reach, in particular when one uses optical
traps rather than magnetic traps \cite{OPT_TRAP}. In this strong
confining regime the motion of the particles along $z$ is
frozen in the ground state of the strong harmonic potential:
\begin{equation}
\phi(x,y,z)\simeq \psi(x,y)
\left({m\omega_z\over \pi\hbar}\right)^{1/4}e^{-m\omega_z z^2/2\hbar}
\label{eq:frozen}
\end{equation}
and only the dependence of the wavefunction $\psi$ in the
$x-y$ plane remains to be determined.
Inserting Eq.(\ref{eq:frozen}) in the energy functional 
Eq.(\ref{eq:ener3d}) we get the corresponding energy functional
for $\psi$ to be minimized, dropping the constant term ${1\over 2}
\hbar\omega_z$:
\begin{equation}
E[\psi,\psi^*] = \int d^2\vec{r}\
{\psi^*(\vec{r}\,)\left[ {\cal H}_\perp-\Omega L_z
\right]\psi(\vec{r}\,)
\over \langle\psi|\psi\rangle} +{1\over 2}
{Ng |\psi|^4 \over \langle\psi|\psi\rangle^2}.
\label{eq:ener2d}
\end{equation}
This 2D energy functional gives the energy
per particle in the condensate measured from the zero-point
energy along $z$.
The 2D Hamiltonian  is
\begin{equation}
{\cal H}_\perp=-{\hbar^2\over 2m}\Delta_{x,y} + {1\over 2} \sum_{\alpha=x,y}
m\omega_\alpha^2 r_\alpha^2.
\end{equation}
The trap is now rotated around the $z$ axis at the angular velocity $\Omega$.
The interaction term $|\psi|^4$ involves an effective 2D coupling
constant between the atoms 
\cite{MAXIM} 
\begin{equation}
g=g_{3D}\left({m\omega_z\over 2\pi\hbar}\right)^{1/2}.
\end{equation}
Most of the results of the paper are dealing with the 2D energy functional;
a numerical result for a local minimum of
the full 3D energy functional will be given in the
section \ref{sec:3d}. We concentrate on the so-called Thomas-Fermi
regime, where the interaction energy per particle is much larger than
$\hbar\omega_{x,y}$. The opposite regime has already been
studied in \cite{ROKHSAR}.

\section{Local minima of energy with vortices} \label{sec:prop}

In this section we briefly discuss the general problem
of minimizing energy functionals of the type Eq.(\ref{eq:ener2d}).
We present the numerical algorithm that we have used and we give
numerical results for the 2D problem.

\subsection{A numerical algorithm to find local minima}

The algorithm in our numerical calculations is commonly used
in the literature to minize energy functionals $E[\psi,\psi^*]$
of the form Eq.(\ref{eq:ener2d}). The intuitive idea is to
start from a random $\psi$ and move it opposite to the local 
gradient of $E[\psi,\psi^*]$ that is along the 
local downhill slope of the energy.
Numerically this is implemented by an evolution of $\psi$ 
parametrized by a fictitious time $\tau$:
\begin{equation}
-{d\over d\tau}\psi = {\delta E\over\delta\psi^*}[\psi,\psi^*].
\end{equation}
Assuming a $\psi$ normalized to unity we get
the following equation of motion for $\psi$:
\begin{equation}
-{d\over d\tau}\psi = [{\cal H}_\perp -\Omega L_z +Ng|\psi|^2-\mu(\tau)]\psi
\end{equation}
that is a non-linear Schr\"odinger equation in complex time $t=-i\tau$.
The quantity $\mu$ appearing in this equation can be expressed explicitly
in terms of a functional of $\psi,\psi^*$; it ensures
that the norm of $\psi$ does not evolve with $\tau$.

This equation is, for $\Omega=0$, standardly solved by a splitting technique, 
propagating during the time step $d\tau$ first with potential energy
in position space (where it is diagonal)
then with kinetic energy in momentum space (where the Laplacian is diagonal).
One goes back and forth between position and momentum space with
Fast Fourier Transforms along $x$ and $y$.
In our case $\Omega\neq 0$ and the Hamiltonian contains
$L_z = x p_y-y p_x$; we have therefore complemented the splitting
scheme by (1) a propagation during $d\tau$ due to $-\hbar\Omega x p_y$
in position space along $x$ and momentum space along $p_y$,
and (2) a similar procedure for the $\hbar\Omega y p_x$ propagation,
that is in momentum space along $x$ and position space along $y$.

One can check that the mean energy of
$\psi$ is a decreasing function of $\tau$:
\begin{equation}
{d\over d\tau}E[\psi,\psi^*] = -2 \int d^2\vec{r}\ 
\left|{\delta E\over\delta\psi^*}\right|^2 \leq 0.
\end{equation}
In the case we consider $E$ remains always positive:  
the scattering length and therefore $g$
are positive so that the interaction term is positive,
and $|\Omega|$ is smaller than the trap frequencies $\omega_{x,y}$
so that the centrifugal potential $-{1\over 2}m\Omega^2r^2$
cannot exceed the trapping potential. Therefore $E$ has to converge 
to a finite value for $\tau\rightarrow\infty$. Asymptotically
$dE/d\tau=0$ and $\psi$ satisfies
${\delta E\over \delta \psi^*}[\psi,\psi^*]=0$, so that
we recover for $\tau=\infty$ the time independent Gross-Pitaevskii
equation:
\begin{equation}
\label{eq:gpe}
\mu\psi = {\cal H}_\perp\psi +Ng|\psi|^2\psi -\Omega L_z \psi
\end{equation}
where $\mu=\mu(\tau=\infty)$ is now the chemical potential of
the gas \cite{REVUE}.

As we have started from a random wavefunction $\psi$, without assuming
any symmetry properties of $\psi$, we expect the trajectory $\psi(\tau)$
to converge as $\tau\rightarrow\infty$ to a local minimum of
the energy functional. We have checked this assumption
by adding a small random wavefunction to $\psi$ and resuming the
evolution in $\tau$; $\psi$ was relaxing to its initial value.
Mathematically the steady state solutions  for the $\tau$
evolution that we find are
stable, which is equivalent to saying
that they are local minima of the energy.
Note
that not all solutions of the Gross-Pitaevskii equation share this
property: the Gross-Pitaevskii equation expresses only the fact
that the energy functional is stationary in $\psi$, which is the case e.g.\
at saddle points of the energy functional (an example is 
a vortex with a charge $|q|>1$, see section \ref{sec:why}).

\subsection{Numerical results in 2D\label{subsec:num2d}}

Applying the algorithm detailed in the previous section 
we present results on local minima of the energy functional
Eq.(\ref{eq:ener2d}) for asymmetric and symmetric traps in the $x-y$
plane.  We characterize the non-axisymmetry of the trap
by $\epsilon$ such that
\begin{eqnarray}
\omega_x &=&\omega/(1+\epsilon) \\
\omega_y &=&\omega (1+\epsilon).
\end{eqnarray}
In Fig.\ref{fig:aniso} we show different local minima configurations
obtained for $\epsilon=0.3$ and a rotation frequency
$\Omega=0.2\omega$; each configuration has been obtained
for different random initial $\psi$'s. The holes observed in the
spatial density correspond to the vortex cores. We have always
found that the phase of $\psi$ changes by $2\pi$  around a vortex core;
we have not found vortices with a charge $\pm q$,
where the integer $q$ is strictly larger than one; this fact will
be explained in the next section. Furthermore the sense of circulation
is the same for all vortices.

To quantify the effect of the non-axisymmetry of the trap we
have plotted in Fig.\ref{fig:eeps} the dependence of energy 
of different vortex configurations on $\omega_x/\omega_y$ for
a fixed $\omega$; we 
measure the energies from $E_{iso}$, the energy of the
zero-vortex solution in the axisymmetric case $\epsilon=0$. 
The zero-vortex
solution exhibits a significant variation of energy with $\epsilon$;
for a non-zero $\epsilon$ the wavefunction $\psi$ develops
a phase proportional to $\Omega$ for weak $\Omega$'s,
which accounts for the energy change as
explained in section \ref{subsec:phase}.
The solutions with vortices experience quasi
the same energy shift as function of $\epsilon$. As only the energy
difference between the various local minima matters we
will from now on only consider the axisymmetric case $\epsilon=0$
to identify the solution with the absolute minimal energy.

Note that the solutions $\psi$ with several vortices
obtained in the limiting case $\epsilon=0$  are not eigenvectors
of $L_z$; this reflects a general property of non-linear equations 
such as the Gross-Pitaevskii equations to have {\it symmetry broken
solutions}; it is explained in \cite{ROKHSAR} how to reconcile
this symmetry breaking with the fact that eigenvectors of the full
$N$-atom Hamiltonian are of well defined angular momentum.

\section{Stability properties of vortices}\label{sec:why}

In this section
we recall that a (normalized) wavefunction $\psi$ such that
$E[\psi,\psi^*]$ has a local minimum in $\psi$, describes
a condensate having all the desired properties of stability,
that is dynamical and thermodynamical stability. We then
show that a vortex centered at $\vec{r}=\vec{0}$ 
with an angular momentum strictly larger
than $\hbar$ is not a local minimum of energy and is therefore
thermodynamically unstable.

\subsection{Stability properties of local minima}
Let us express the fact that $\psi$ corresponds to a local
minimum of the energy. A first condition is that the energy
functional is stationary for $\psi$, that is $\psi$ solves
the Gross-Pitaevskii equation Eq.(\ref{eq:gpe}). To get the
second condition, 
we consider a small variation of
$\psi$,
\begin{equation}
\psi \rightarrow \psi +\delta\psi
\label{eq:dev}
\end{equation}
preserving the normalization of the condensate wavefunction to 
unity: 
\begin{equation} \label{eq:norm}
||\psi+\delta\psi||^2-||\psi||^2 =0= \langle\psi|\delta\psi\rangle +
\langle\delta\psi|\psi\rangle+\langle\delta\psi|\delta\psi\rangle.
\end{equation}
We expand the energy functional $E[\psi,\psi^*]$ in powers of
$\delta\psi$, neglecting terms of order $\delta\psi^3$ or higher.
Using Eq.(\ref{eq:norm}) and Eq.(\ref{eq:gpe}) we find that terms
linear in $\delta\psi$ vanish so that
\begin{equation}
\delta E = {1\over 2} (\langle\delta\psi|,\langle\delta\psi^*|)
{\cal L}_c 
\left(\begin{array}{c}
|\delta\psi\rangle \\ |\delta\psi^*\rangle
\end{array}
\right) + o(\delta\psi^2).
\label{eq:vare}
\end{equation}
We have introduced the operator
\begin{equation} \label{eq:lc}
{\cal L}_c = 
\left(
\begin{array}{cc}
{\cal H}_{GP} + Ng |\psi|^2  & Ng \psi^2  \\
Ng \psi^{*^2} & {\cal H}_{GP}^* + Ng  |\psi|^2 \\
\end{array}
\right)
\end{equation}
and the Gross-Pitaevskii Hamiltonian
\begin{equation}
{\cal H}_{GP} = {\cal H}_\perp + Ng |\psi|^2 -\Omega L_z-\mu.
\end{equation}
The fact that $E$ has a local minimum in $\psi$ imposes that
the Hermitian operator ${\cal L}_c$ be positive. In general
${\cal L}_c$ will be strictly positive apart from the zero energy mode
$(i\psi,-i\psi^*)$ corresponding to an inessential change of the
global phase of $\psi$. We now show that the positivity of ${\cal L}_c$
implies  the stability of the solution $\psi$.

\subsubsection{Dynamical stability}
Consider first the problem of so-called \lq\lq dynamical stability":
to be a physically acceptable condensate wavefunction,
$\psi$ has to be a stable solution of the time dependent Gross-Pitaevskii
equation
\begin{equation}
i\hbar\partial_t\psi = {\cal H}_{GP}\psi
\label{eq:gpet}
\end{equation}
otherwise any small perturbation of $\psi$, e.g.\ the effect of quantum
fluctuations or experimental noise, may lead to an evolution of
$\psi$ far from its initial value. To determine the evolution of
a small deviation $\delta\psi$ as in Eq.(\ref{eq:dev}) we linearize
Eq.(\ref{eq:gpet}):
\begin{equation}
i\hbar\partial_t \left(\begin{array}{c}
|\delta\psi\rangle \\ |\delta\psi^*\rangle
\end{array}
\right)
=
{\cal L}
\left(\begin{array}{c}
|\delta\psi\rangle \\ |\delta\psi^*\rangle
\end{array}
\right)
\end{equation}
where the operator ${\cal L}$ is related to ${\cal L}_c$ by
\begin{equation}
\label{eq:defl}
{\cal L}_c = \left(
\begin{array}{cr}
1 & 0\\
0 & -1
\end{array}
\right)
{\cal L}.
\end{equation}
As $\psi$ is time independent, so is $\cal L$
and dynamical stability is equivalent to the requirement
that the eigenvalues of $\cal L$ have all a negative or vanishing
imaginary part. As we now show the positivity of ${\cal L}_c$
leads to a purely real spectrum for $\cal L$. Consider
an eigenvector $(u,v)$ of $\cal L$ with the eigenvalue $\varepsilon$.
Contracting Eq.(\ref{eq:defl}) between
the ket $(|u\rangle,|v\rangle)$ and the bra $(\langle u|,\langle v|)$
we get 
\begin{equation}
\varepsilon \left[\langle u|u\rangle-\langle v|v\rangle\right]=
(\langle u|,\langle v|){\cal L}_c \left(
\begin{array}{c}
|u\rangle \\
|v\rangle
\end{array}
\right).
\label{eq:sand}
\end{equation}
Note that the matrix element of ${\cal L}_c$ is real positive as
${\cal L}_c$ is a positive hermitian operator. We now face two
possible cases for the real quantity $\langle u|u\rangle-\langle v|v\rangle$:
\begin{itemize}
\item $\langle u|u\rangle-\langle v|v\rangle= 0$. In this case 
${\cal L}_c$ has a vanishing expectation value in
$(|u\rangle,|v\rangle)$; as ${\cal L}_c$ is positive
$(|u\rangle,|v\rangle)$ has to be an eigenvector 
of ${\cal L}_c$ with the eigenvalue zero; from Eq.(\ref{eq:defl})
and the fact that $\left(\begin{array}{cr} 1 & 0 \\
0 & -1\end{array}\right)$ is invertible we find that $(|u\rangle,|v\rangle)$
is also an eigenvalue of $\cal L$ with the eigenvalue 0, so that $\varepsilon=0$
is a real number. 
\item $\langle u|u\rangle-\langle v|v\rangle > 0$:
we get $\varepsilon$ as the ratio
of two real numbers, so that $\varepsilon$ is real.
\end{itemize}

\subsubsection{Thermodynamical stability}
A second criterion of stability is the so-called \lq\lq thermodynamical"
stability. 
For zero temperature,
this condition can be formulated in the Bogoliubov
approach \cite{REVUE}, where the particles out of the condensate,
which always exist because of the interactions,
are described by a set of uncoupled harmonic oscillators with
frequencies 
$\varepsilon \mbox{sign}\left[ \langle u|u\rangle-\langle v|v\rangle\right]
/\hbar$, where $(u,v)$ is an eigenvector of
$\cal L$ with the eigenvalue $\varepsilon$. 
In order for a thermal equilibrium to exist for these
oscillators, their frequencies should be strictly positive, which is the
case here in virtue of Eq.(\ref{eq:sand}) \cite{APART}.
If a mode with a negative frequency
were present thermalization by collisions would transfer particles from
the condensate $\psi$ to this mode, leading to a possible evolution of
the system far from the initial state $\psi$ \cite{ROKSTAB}.

What happens for solutions $\psi$ of the Gross-Pitaevskii equations
that are not local minima of energy ? The operator ${\cal L}_c$
has at least an eigenvector with a strictly negative eigenvalue.
In this case one cannot have thermodynamical stability, that is one
cannot have $\varepsilon\left[ \langle u|u\rangle-\langle v|v\rangle\right] 
>0$ for all modes \cite{APART}. From the non-positivity of ${\cal L}_c$
one cannot however distinguish between a simple thermodynamically
instability or a more dramatic dynamical instability.

\subsection{Why not a vortex of angular momentum larger than $\hbar$ ?}
\label{sec:ana}

For simplicity we consider only a single vortex
in the center of an axi-symmetric trap. 
We show that vortices with a change of phase of $2q\pi$ 
are not local minima of energy, that is are (at least thermodynamically)
unstable.
We have found numerically
a solution of the Gross-Pitaevskii equation Eq.(\ref{eq:gpe}) by an 
evolution in complex time, starting from 
a wavefunction $\psi$ with an angular momentum $q\hbar$ along $z$,
as already done in \cite{CRIT}; our solution of the Gross-Pitaevskii
equation with imposed symmetry
is a local mimimum of energy in the subspace of 
functions with angular momentum
$q\hbar$ along $z$, 
but not necessarily a local minimum in the whole functional
space, as we will see for $|q|>1$.
In the Thomas-Fermi regime
$\mu\gg\hbar\omega$ we find that the
solutions can be well reproduced by a variational Ansatz of
the form
\begin{equation}
\psi(x,y) = e^{iq\theta} \left[\tanh\kappa_q r\right]^{|q|}
\left({\tilde{\mu}-{1\over 2}m\omega^2r^2\over Ng}\right)^{1/2}
\label{eq:ans1}
\end{equation}
where $\theta$ is the polar angle in the $x-y$ plane and
where $\tilde{\mu}$, the chemical potential in the lab frame
\begin{equation}
\tilde{\mu} = \mu + q\hbar\Omega
\end{equation}
does not depend on $\Omega$. In this Ansatz the vortex core
is accounted for by $\tanh^{|q|}$, a function that vanishes  as $r^{|q|}$
in zero as it should, and the condensate density outside the core coincides
with the Thomas-Fermi approximation commonly used
for the zero-vortex solution \cite{REVUE}.
We calculate the mean energy 
Eq.(\ref{eq:ener2d}) of the
variational Ansatz and we minimize it with
respect to the variational parameter $\kappa_q$; we get
\begin{equation}
\kappa_q =\left[\tilde{\mu} m\over \hbar^2 \right]^{1/2} c_q 
\end{equation}
where 
\begin{equation}
c_q^2 =  {2\over q^2} \int_0^{+\infty}du\ u\left(\tanh^{2|q|}(u)-1\right)^2
\end{equation}
is a number ($c_1=0.7687, c_2=0.5349,\ldots$).

In order for the vortex of charge $q$ to be a local minimum of energy, the
operator ${\cal L}_c$ of Eq.(\ref{eq:lc}) has to be positive. This
implies that the operator on the first line, first column of 
${\cal L}_c$, the so-called Hartree-Fock Hamiltonian,
be positive:
\begin{equation}
{\cal H}_{HF} = {\cal H}_{\perp} + 2 Ng |\psi|^2 -\tilde{\mu}+q\hbar\Omega
-\Omega L_z \geq 0.
\end{equation}
To show that this is not the case it is sufficient to find
a wavefunction $f(x,y)$ leading to a negative expectation 
value for ${\cal H}_{HF}$. As the potential appearing in
${\cal H}_{HF}$ has a dip at $r=0$ we have taken $f$ of a form
localized around $r=0$:
\begin{equation}
f(x,y) = {1\over 
\cosh \left[\gamma\left({\tilde{\mu}m\over \hbar^2}\right)^{1/2}
r\right]}
\end{equation}
where $\gamma$ is adjusted to minimize the expectation value.
For e.g.\ $q=2$ we take $\gamma=1$ leading to
\begin{equation}
{\langle f|H_{HF}|f \rangle\over \langle f|f \rangle}
\simeq -0.407 \tilde{\mu} + 2\hbar\Omega.
\end{equation}
As $\Omega <\omega\ll \mu$ this quantity is negative. A similar conclusion
is obtained for $q>2$.

We have also performed a numerical experiment, 
evolving $\psi$ in complex time starting from 
Eq.(\ref{eq:ans1}); we find that the vortex $q=2$ splits in two
vortices $q=+1$ symmetrically dispatched
\cite{ROKSTAB}. A numerical diagonalization of $\cal L$ shows
that the vortex $q=2$ is alternatively dynamically and thermodynamically
unstable when one increases $\tilde{\mu}/\hbar\omega$ \cite{BIGELOW}.

\section{How to detect the presence of vortices ?} \label{sec:how}

Several signatures of the presence of vortices
have been proposed in the literature.
A first possibility is a measurement of
the excitation spectrum as studied in
\cite{STAB}. Another idea is to measure the second order correlation
function of the atomic field \cite{INTER}.

A third signature of the presence of vortices is also the holes 
in the density due to the vortex cores. As the size of the vortex
core in the Thomas-Fermi regime 
is too small to be observed in situ by optical imaging techniques,
we suggest to switch off the trapping potential and let the cloud
expand; as we now check the size of the cloud and the size of the
vortex cores are magnified by the same factor in the expansion,
so that the cores become observable.

To study the expansion of the gas when the trap in the 
$x-y$ plane is switched off, the confinement along $z$ being kept
constant,
one has to solve a 2D time dependent Gross-Pitaevskii equation.
In this section the trap is axi-symmetry with a time
dependent frequency $\omega(t)$. We consider the evolution
in the laboratory frame, as the detection is performed in
this frame:
\begin{equation}
i\hbar\partial_t \psi_{lab} = \left[-{\hbar^2\over 2m} \Delta
+{1\over 2} m\omega^2(t)r^2+Ng|\psi_{lab}|^2\right]\psi_{lab}.
\end{equation}
As shown in \cite{SCAL1,SCAL2} the effect of the
time dependence of $\omega(t)$ can be absorbed by a scaling
and gauge transform of the wavefunction:
\begin{equation}
\label{eq:psit}
\psi_{lab}(\vec{r},t) = {1\over\lambda(t)} 
e^{imr^2\dot{\lambda}/2\hbar\lambda}
\tilde{\psi}(\vec{r}/\lambda(t),t)
\end{equation}
where $\omega(0)$ is the oscillation frequency before opening
the trap; the scaling parameter solves:
\begin{equation}
\ddot{\lambda} = {\omega^2(0)\over \lambda^3}-\omega^2(t)\lambda 
\end{equation}
with initial conditions $\lambda(0)= 1, \dot{\lambda}(0) = 0$; 
if the trap in the $x-y$ plane is abruptly switched off at $t=0^+$ 
the scaling parameter is given by
\begin{equation}
\lambda(t) = \sqrt{1+\omega^2(0)t^2}.
\end{equation}
Introducing the renormalized time $\tau$ given by
\begin{equation}
{dt\over\lambda^2(t)} = d\tau
\end{equation}
we find that $\tilde{\psi}$ solves the same equation
as $\psi_{lab}$ with a {\it constant} trap frequency equal to 
$\omega(0)$:
\begin{equation}
i\hbar\partial_{\tau}\tilde{\psi}=  \left[-{\hbar^2\over 2m} \Delta
+{1\over 2} m\omega^2(0)r^2+Ng|\tilde{\psi}|^2\right]\tilde{\psi}.
\end{equation}
As $\psi_{lab}$ rotates in the trap at the frequency $\Omega$ in the lab
frame, so does $\tilde{\psi}$ in terms of the renormalized
time $\tau$. In the limit of $t\rightarrow\infty$,
$\tau$ tends to a finite value $\tau_{max}$, so that 
$\tilde{\psi}$ is rotated by a finite angle during the ballistic
expansion:
\begin{equation}
\Omega\tau_{max}=\Omega\int_0^{\infty} {dt\over\lambda^2(t)}= {\pi\over 2}
{\Omega\over \omega(0)}.
\end{equation}
Therefore $\psi_{lab}$ rotates with respect to its value
when the trap is switched off and its size is magnified by $\lambda(t)$.

A fourth possibility, giving direct access to the phase of the vortex,
is to measure the interference fringes between a condensate with
vorticity and a reference condensate with no vortex.
We study this possibility as an application of the scaling solution \cite{PRIV}.
The condensate 2 has one or several vortices and
is originally centered at $\vec{r}=0$, the condensate 1
has no vortex and is centered initially at $\vec{r}=\vec{d}$.
After ballistic expansion of the condensates the resulting density
can be written:
\begin{equation}
\rho_{1+2}=|\rho_1^{1/2} e^{im(\vec{r}-\vec{d})^2/2\hbar t}
+\rho_2^{1/2} e^{imr^2/2\hbar t}e^{iS(\vec{r})}e^{i\gamma(t)}|^2
\end{equation}
where $\rho_{1}$ and $\rho_{2}$ are the densities of the condensates
1 and 2 respectively,
$S$ is the phase due to vorticity of the condensate 2, $\gamma(t)$
is a relative phase depending only on time; to obtain
the phase terms quadratic in $\vec{r}$ we have used Eq.(\ref{eq:psit})
with the asymptotic value 
$\dot{\lambda}/\lambda\simeq 1/t$ for $t\rightarrow\infty$.
We have plotted an example of interference fringes with
two vortices in Fig.\ref{fig:interf}.

The above scaling result is exact only for axi-symmetry 2D traps.
For a non axi-symmetric traps and for 3D situations where
the confinement along $z$ is not strong, it has been shown that
approximate scaling solutions exist in the absence of a vortex
\cite{SCAL1,SCAL2}; in presence of vortices  we have integrated
numerically the time dependent Gross-Pitaevskii equation
and found that the density experiences an approximate scaling,
the magnification of the vortex being slightly larger
than the one of the cloud (see also \cite{PETHICK}).

\section{Intuitive variational calculation} \label{sec:intuitive}
To get a better understanding of the numerical results
we now proceed to an intuitive Ansatz for the wavefunction 
with several vortices. It coincides very well with the numerical
results and allows an easy construction of the minimal energy configurations
with vortices. It gives a physical understanding 
of the stability conditions and of the structure of the solutions:
a set of $n$ vortices is 
equivalent to a gas of interacting particles in presence
of an external potential adjusted by the rotation frequency
of the trap.
We restrict to the case of an axi-symmetric
trap, a good approximation for weak ($<10\%$) non-axisymmetries
(see section \ref{subsec:num2d}).

\subsection{Ansatz for the density}

To construct the Ansatz we split $\psi$ in a modulus and a phase:
\begin{equation}
\psi(x,y)= |\psi|e^{iS}.
\label{split}
\end{equation}
In the Thomas-Fermi regime, 
the modulus in presence of $n$ vortices appears as a slowly
varying envelope given by the Thomas-Fermi approximation used
in the 0-vortex case:
\begin{equation}
\psi_{slow} = \left[{\mu-{1\over 2}m\omega^2r^2}\over Ng\right]^{1/2}
\end{equation}
with narrow holes digged by the vortices with charge $q=\pm1$, represented by 
$\tanh$ functions of adjustable widths and with zeros at 
adjustable positions:
\begin{equation}
|\psi|=\psi_{slow}\times 
\prod_{k=1}^{n} \tanh[\kappa_k|\vec{r}-\vec{\alpha}_k R|].
\label{eq:ans}
\end{equation}
The positions of the vortex cores $\vec{\alpha_k}$ are expressed
in units of the Thomas-Fermi radius $R$ of the condensate:
\begin{equation}
R=\sqrt{2\mu\over m\omega^2}.
\end{equation}
From section \ref{sec:ana} we expect as typical values for the inverse
width of the vortex cores
$\kappa_k \simeq (m\mu/\hbar^2)^{1/2}$.
The chemical potential is not an independent variable but is expressed
as a function of the other parameters from the normalization
condition $\langle\psi|\psi\rangle=1$; neglecting overlap
integrals between the holes we get
\begin{equation}
\mu = \mu_0 \left[1+2\sum_{k=1}^{n} (1-\alpha_k^2){\ln 2\over (\kappa_k R)^2}
+O({1\over (\kappa R)^4})\right]
\end{equation}
where $1/(\kappa R)^4 \sim (\hbar\omega/\mu)^4 \ll 1$ and
where $\mu_0$ is the Thomas-Fermi approximation for the condensate
chemical potential without vortices:
\begin{equation}
\mu_0 = \left({m\omega^2 N g\over \pi}\right)^{1/2}.
\end{equation}

\subsection{The phase}\label{subsec:phase}
The general form of the phase of $\psi$ in
Eq.(\ref{split}) in presence
of $n$ vortices is:
\begin{equation} \label{SXY}
S(x,y) = \sum_{k=1}^{n} q_k\theta_k +S_0(x,y)
\end{equation}
where the integer $q_k=\pm 1$ is the vortex charge
(that is the angular momentum (over $\hbar$)
of the vortex $k$ with respect to its core
axis), $\theta_k$ is the polar angle of a
system of Cartesian coordinates $(X,Y)$ centered on the vortex core
and $S_0$ is the single-valued part of the phase. The function $S_0$ 
can in principle be determined from the modulus of $\psi$ from
the continuity equation:
\begin{equation}
\mbox{div}[|\psi|^2\vec{v}\,]=0.
\end{equation}
The local velocity field $\vec{v}$ is related to the phase
$S$ by
\begin{equation}
\vec{v}={\hbar\over m}\vec{\nabla} S -\vec{\Omega}\wedge\vec{r}.
\end{equation}
This expression is derived from the relation between the
velocity operator and the momentum operator in the
rotating frame, $\hat{\vec{v}}= \hat{\vec{p}}/m-\vec{\Omega}\wedge
\hat{\vec{r}}$.
Expanding the continuity equation we obtain
\begin{equation}
|\psi|^2\Delta S +\vec{\nabla}|\psi|^2\cdot\vec{\nabla} S-{m\Omega\over\hbar}
[x\partial_y-y\partial_x]|\psi|^2 = 0.
\label{eq:cont}
\end{equation}
This can be turned into an equation for the single-valued
part $S_0$ of the phase; because the density $|\psi|^2$ in a trap
vanishes at the border of the condensate $S_0$ is uniquely determined
(up to a constant) by the resulting equation (see Appendix); 
this is to be contrasted to the case of superfluid helium in
a container, where  
the flux, not the density,
vanishes at the border, which requires a boundary condition on
the gradient of the phase.

Eq.(\ref{eq:cont}) can be solved for a non-axisymmetric trap in the
absence of vortices. The solution is given by
\begin{equation}
\label{eq:fet}
S(x,y)= - {m\Omega\over \hbar} {\omega_x^2-\omega_y^2\over
\omega_x^2+\omega_y^2}xy
\end{equation}
which leads to a change in the energy per particle 
\begin{equation}
\delta E = -{1\over 6}\mu_{TF}
\Omega^2{(\omega_x^2-\omega_y^2)^2\over (\omega_x^2
+\omega_y^2)\omega_x^2\omega_y^2}
\end{equation}
where $\mu_{TF}$ is the Thomas-Fermi approximation for the
chemical potential for $\Omega=0$, $\mu_{TF} = 
(m\omega_x\omega_y N g/\pi)^{1/2}$ \cite{FETTER}. 
As can be seen in Fig.\ref{fig:eeps} 
this prediction is in good agreement with our
numerical results.

In presence of vortices the equation for $S$ is more difficult to solve
analytically. From now on we consider the
case of an axi-symmetric trap, as the energy ordering of the
vortices solutions is not affected for weak
($<10\%$) non-axisymmetries (see section \ref{subsec:num2d}). 
For a single vortex at the center of the trap one can see that $S_0=0$
solves Eq.(\ref{eq:cont}).
From the spatial dependence of the
phase obtained numerically (section \ref{subsec:num2d})
for a displaced vortex or several vortices
we have identified the following heuristic Ansatz, obtained in
setting $\omega_x=\omega_y=\omega$ in Eq.(\ref{eq:fet}):
\begin{equation}
S_0(x,y)\equiv 0
\end{equation}
that we will use in the remaining part of the section.
\subsection{Further approximations for the mean energy}
In the calculation of the mean energy, 
we make some further approximations 
in the spirit of the Ansatz Eq.(\ref{eq:ans}). The reader
not interested by these more technical considerations can 
proceed to the next subsection.

The kinetic energy involves an integral of the gradient squared
of the wavefunction:
\begin{equation}
|\vec{\nabla}\psi|^2 = |\psi|^2\left[(\vec{\nabla}\ln|\psi|)^2 + (\vec{\nabla} S)^2\right].
\label{eq:grad}
\end{equation}
For the gradient of the modulus of $\psi$ we neglect the variation of
the slow envelope $\psi_{slow}$:
\begin{equation}
\vec{\nabla} \ln|\psi| \simeq \sum_{k=1}^{n} \kappa_k{\tanh'\over
\tanh}[\kappa_k|\vec{r}-\vec{\alpha_k}R|]\left(\vec{e}_r\right)_k
\label{eq:grad2}
\end{equation}
where $\left(\vec{e}_r\right)_k = 
(\vec{r}-\vec{\alpha_k}R)/|\vec{r}-\vec{\alpha_k}R|$.
The terms in this sum are peaked around the vortex cores; assuming a separation
between the vortex cores much larger than their width, we neglect all the 
crossed terms in the square of Eq.(\ref{eq:grad2}).
Consider now the second term in Eq.(\ref{eq:grad}). The gradient squared
of $S$ involves diagonal terms $(\vec{\nabla}\theta_k)^2$ and
non-diagonal terms $\vec{\nabla}\theta_k\cdot\vec{\nabla}\theta_{k'}$;
the modulus squared of $\psi$ involves holes with a density varying
as $1-\tanh^2=\sech^2$. In the following we keep 
the $\sech^2$ for the vortex $k$ only if it is multiplied by 
$(\vec{\nabla}\theta_k)^2$, a quantity diverging in the center
of the core; the other terms lead to converging integrals
smaller by a factor $(\mu/\hbar\omega)^2$, which is the inverse
surface of a vortex core ($\int d^3\vec{r}\ |\psi_{slow}|^2\sech^2\kappa r
\propto 1/(\kappa R)^2$). This finally leads to
\begin{eqnarray}
E_{kin}  \simeq {\hbar^2\over 2m}\int d^2\vec{r}\
|\psi_{slow}|^2 &&
\left[\sum_{k=1}^{n}
\tanh^2[\kappa_k|\vec{r}-\vec{\alpha_k}R|] (\vec{\nabla}\theta_k)^2
+\kappa_k^2
\left(\tanh'[\kappa_k|\vec{r}-\vec{\alpha_k}R|]\right)^2 
\right.\nonumber \\
&+&
\left. \sum_{k=1}^{n}\sum_{k'\neq k} \vec{\nabla}\theta_k\cdot\vec{\nabla}\theta_{k'}
\right]
\end{eqnarray}

In the same spirit we simplify the contribution of $-\Omega L_z$ to the
energy:
\begin{eqnarray}
E_{rot}& =& -\hbar\vec{\Omega}\cdot\int d^2\vec{r}\ |\psi|^2
\vec{r}\wedge\vec{\nabla}S\\
&\simeq& -\hbar\vec{\Omega}\cdot\int d^2\vec{r}\ |\psi_{slow}|^2
\vec{r}\wedge\sum_{k=1}^{n}\vec{\nabla}\theta_k.
\end{eqnarray}
In the potential energy
\begin{equation}
E_{pot} = \int d^2\vec{r}\ |\psi|^2\left[ {1\over 2}m\omega^2 r^2
+{1\over 2} N g |\psi|^2\right]
\end{equation}
we will neglect in $|\psi|^4$ products of $\sech^2$ coming from
different vortex cores.

\subsection{A more physical form of the mean energy}

After the approximations detailed in the previous subsection
the mean energy in presence of $n$ vortices
is a sum of one-vortex self energies 
and binary interaction energies between the vortices:
\begin{equation}
E={2\over 3}\mu_0 +\sum_{k=1}^{n} W(\vec{\alpha_k},\kappa_k)
+{1\over 2}\sum_{k=1}^{n}\sum_{k'\neq k} V(\vec{\alpha_k},\vec{\alpha_{k'}}).
\label{suggest}
\end{equation}
Eq.(\ref{suggest}) allows to interpret a system with $n$ vortices as
a gas of \lq\lq particles"
with binary interactions; the form of the interaction potential
obtained here is valid for a separation between the 
\lq\lq particles" larger than the size $1/\kappa$ of the vortex cores.
The vortex self-energy is
\begin{eqnarray} \label{SELFFORM}
W(\vec{\alpha},\kappa) &=&
{(\hbar\omega)^2\over \mu_0}\left\{
{1\over 2} +(\alpha^2-1)\left[C-\ln\kappa R
-{1\over 2}\ln(1-\alpha^2)\right]\right\}\nonumber \\
&&-q\hbar\Omega \left[(1-\alpha^2)^2 \right]\nonumber \\ 
&&+ {\mu_0\over 3}\left[(4\ln 2-1){(1-\alpha^2)^2\over (\kappa R)^2}
\right]
\end{eqnarray}
where $C = 0.495063$. The lines in Eq.(\ref{SELFFORM}) correspond
successively to $E_{kin}$, $E_{rot}$ and $E_{pot}$.
This can be seen as an effective potential for the vortices.
One can check that the part of $W$ independent of $\Omega$ expells
the vortex core from the trap center, whereas  the
part proportional to $\Omega$ provides a confinement of the vortex core
(see the following subsection).

The vortex interaction potential is given by
\begin{equation} \label{INTFORM}
{1\over 2}V(\vec{\alpha},\vec{\beta})
= {\hbar^2\over 2m} q_{\alpha} q_{\beta}
\int d^2\vec{r}\ |\psi_{slow}|^2
\vec{\nabla}\theta_{\vec{\alpha}R}\cdot \vec{\nabla}\theta_{\vec{\beta}R}.
\end{equation}
This interaction term is equivalent to the one found in the homogeneous
case and describes a repulsive interaction for vortices turning in
the same direction ($q_{\alpha} q_{\beta} >0$ )
and is attractive for vortices with opposite charges
\cite{HESS}.
An attractive interaction will lead to the coalescence 
and consequently annihilation of vortices with opposite charges.
Therefore we find in stationary systems always vortices with equal charges.

As the interaction potential $V(\vec{\alpha},\vec{\beta})$ 
does not depend on the parameters $\kappa$ we 
can optimize separately
the self-energy part with respect to $\kappa$ and
find
\begin{equation}
(\kappa R)^2 = \xi ^2 (1-\alpha^2)\left({\mu_0\over\hbar\omega}\right)^2
\end{equation}
where $\xi = \left[ {2\over 3}(4\ln 2-1)\right]^{1/2} \simeq 1.08707$.
By rewriting the above equation as
\begin{equation}
{\hbar ^2\kappa^2\over m} = {1\over 2}\xi^2\left[\mu-{1\over 2}m\omega^2
(\alpha R)^2\right]
\end{equation}
we find that $\kappa^2$ is proportional to the local chemical potential at
the position $\alpha R$ of the vortex core.
We finally get the explicit form for the self energy as
\begin{equation} \label{SELF}
W(\vec{\alpha}) = {(\hbar\omega)^2\over\mu_0}
\left\{ {1\over 2}+ (1-\alpha^2)\left[ {2\ln 2+1\over 3}+\ln
{\nu\mu_0\over\hbar\omega}+\ln(1-\alpha^2)
-q{\Omega\mu_0\over\hbar\omega^2}(1-\alpha^2)\right]
 \right\}
\end{equation}
where $\nu=0.49312$.

\subsection{ Case of a single vortex: critical frequencies}

In Fig.\ref{fig:self} we have plotted the self-energy of a vortex
as a function of the displacement of the core from
the trap center, for different values of the rotation 
frequency $\Omega$. 
The analytical prediction coincides very well with the numerical
value \cite{COMMENT}.

For $\Omega=0$ the position of the vortex at the trap center
gives an energy maximum.
For $\Omega >0$ the rotation of the trap provides an effective 
confinement of the vortex core at the center of the trap
for positive charges $q$
(see the term proportional to $\Omega$ in Eq.(\ref{SELF})); from
now one we therefore take all the charges $q_k$ to be equal to $+1$.
For a large enough $\Omega$ we reach a situation where a vortex at the
trap center corresponds to a local energy minimum, by further increasing 
$\Omega$ the vortex state at the trap center becomes a global minimum with
energy less than the condensate without vortex.

The above suggests that we have to distinguish
two critical rotation frequencies:
The first one defines the frequency $\Omega_{stab}$
above which the vortex is a local minimum  of energy.
Above the frequency $\Omega_c$ the single vortex solution has an energy
lower than the condensate without vortex.
We calculate $\Omega_{stab}$ from the condition
$d ^2 W/d \alpha^2 =0$ at $\alpha=0$ and $\Omega_c$ from the condition
$W=0$ at $\alpha=0$:
\begin{eqnarray}
\Omega_{c} &=& 
{\hbar\omega^2\over \mu_0} \ln\left[{C'\mu_0\over\hbar\omega}
\right] \\
\Omega_{stab} &=& 
{\hbar\omega^2\over 2\mu_0} \ln\left[{C'e^{1/2}\mu_0\over\hbar\omega}
\right] 
\end{eqnarray}
where $C'= e^{(2\ln 2+1)/3+1/2}\nu \simeq 1.8011$.
As we are in the regime $\mu_0\gg\hbar\omega$ $\Omega_c$
is approximately twice $\Omega_{stab}$ \cite{FETTER2}.
Our prediction for $\Omega_c$ scale as $(\log\mu_0)/\mu_0$ 
as in \cite{CRIT}, with a coefficient $C'$ leading to better agreement
with the numerics.

\subsection{Case of several vortices}

By integrating Eq.(\ref{INTFORM}) we get an explicit form for the
vortex interaction potential for vortices with equal charges:
\begin{eqnarray} \label{INTER}
V(\vec{\alpha},\vec{\beta}) =
{(\hbar\omega)^2\over\mu_0}
&&\left\{ \alpha^2+\beta^2-1-|\vec{\alpha}\wedge\vec{\beta}|\mbox{atan}\,
\left[{|\vec{\alpha}\wedge\vec{\beta}| \over (1-\vec{\alpha}\cdot\vec{\beta})}
\right] \right. \nonumber \\
&+&\left. {1\over 2} (1-\vec{\alpha}\cdot\vec{\beta})
\log\left[{1-2\vec{\alpha}\cdot\vec{\beta}+\alpha^2\beta^2\over
|\vec{\alpha}-\vec{\beta}|^4}\right] \right\}
\end{eqnarray}
At short distances between the two vortex cores the logarithmic term
in the above expression dominates, leading to a repulsive potential
$\sim -2(1-\vec{\alpha}\cdot\vec{\beta})
\log|\vec{\alpha}-\vec{\beta}|(\hbar\omega)^2/\mu_0$.
In Fig.\ref{fig:inter} we plot the interaction energy between a vortex at the center of the
trap and one of equal charge displaced by $\alpha R$;
the interaction is purely repulsive. A conclusion which essentially holds 
as well for arbitrary vortex positions.
In Fig.\ref{fig:deux} we show the total (interaction $+$ self-energy)
for two vortices symmetrically displaced from the trap center,
as function of the displacement; the analytical prediction
coincides again very well with the numerical results \cite{COMMENT}.
To obtain the equilibrium distance between the two vortex
cores one minimizes the total energy  over $\alpha$ in
Fig.\ref{fig:deux}.

To get the minimal energy
configurations as function of the rotation frequency of the trap,
we minimize our analytical prediction for the energy 
over the positions of the $n=1,2,\ldots$ vortex cores. The result
is shown in Fig.\ref{fig:diag}. Each curve corresponds
to a fixed value of $n$; it starts at $\Omega=\Omega_{stab}(n)$ 
(for $\Omega <\Omega_{stab}(n)$ there is no local minima
of energy with $n$ vortices); it becomes the global energy minimum
for $\Omega=\Omega_c(n)$. We have plotted these two critical
frequencies as function of $n$ in Fig.\ref{fig:crit}.

We have also given numerical results (circles) in Fig.\ref{fig:diag}. 
Even if there is good agreement between analytical and numerical results,
we still need a numerical calculation to check the stability
of the solutions; our simple analytical Ansatz
is indeed not sufficient to predict the destabilization of
a given vortex configuration at high $\Omega$, a phenomenon
studied with a numerical calculation of the Bogoliubov
spectrum for a single vortex in \cite{ISO}.

For a fixed value of the number of vortices $n$ there
may exist local minima of energy, in addition to the global minimum
plotted in Fig.\ref{fig:diag}, a situation
known from superfluid helium \cite{DONNELLY}. E.g.\ for $n=6$ (see
Fig.\ref{fig:n=6})
the global minimum of energy
is given by a configuration with six vortex cores
on a circle; there exists also a local minimum of energy with
one vortex core at the center of the trap and five vortex cores
on a circle. The energy difference per particle between
the two configurations is very small, $\delta E\simeq 0.002
\hbar\omega$ for the parameters
of the figure and probably beyond the accuracy of our variational
Ansatz. For relatively large rotation
frequencies $\Omega$ one can find local minima of energy configurations
with many vortices (see \cite{DONNELLY}
for superfluid helium); we plot two configurations with
18 vortices in Fig.\ref{fig:n=18}, with an energy
difference $\delta E=0.0034\hbar\omega$.

In estimating the physical relevance of these energy
differences one should keep in mind that $N\delta E$ matters,
rather than $\delta E$, where $N$ is the number of
particles in the condensate:
e.g.\ at a finite temperature $T$ the ground energy configuration
is statistically favored as compared to the metastable
one when $N\delta E\gg k_B T$.

\section{Vortices in a 3D configuration} \label{sec:3d}

We have 
extended the numerical calculation to the case of a 3D cigar-shaped
trap, that is 
with a confinement weaker along the rotation axis than
in the $x-y$ plane. Even in this case
rotation of the trap can stabilize the vortex.
We show in Fig.\ref{fig:3d} density cuts of a solution with
5 vortices; the vortex cores are almost straight lines in the
considered Thomas-Fermi regime, except at vicinity of the borders of 
the condensate. As in section \ref{sec:intuitive} the core
diameter is determined by the local chemical
potential in the gas. 

This suggests that our 2D Ansatz (section
\ref{sec:intuitive}) can be generalized to 3D situations,
with $\vec{\alpha}_k$ and $\kappa_k$ depending on $z$.

\section{Conclusion and perspectives}

We have presented in this paper
an efficient numerical algorithm and a heuristic
variational Ansatz to determine the local minima energy configurations
for a Bose-Einstein condensate strongly confined along $z$
and subject to a rotating harmonic trap in the $x-y$ plane.

Our results can be used as a first step towards finite temperature 
calculations. Interesting problems are e.g.\ the
critical temperature for the vortex formation and
the Magnus forces induced by the non-condensed particles
on the vortex core \cite{SONIN}.

{\bf Acknowledgement:} We acknowledge useful discussions with
Sandro Stringari and Dan Rokhsar. We thank J.\ Dalibard for
useful comments on the manuscript.
We thank the ITP at Santa Barbara for its hospitality and
the NSF for support under grant No. PHY94-07194.
This work was partially supported by
the TMR Network 
\lq\lq Coherent Matter Wave Interactions", FMRX-CT96-0002.
{$^*$ L. K. B.
is an unit\'e de recherche de l'Ecole Normale Sup\'erieure et de l'Universit\'e
Pierre et Marie Curie, associ\'ee au CNRS.}

\appendix
\section{Uniqueness of the phase from the continuity equation in a trap}
Consider two solutions $S_1$ and $S_2$ of the continuity
equation:
\begin{equation}
\mbox{div}[|\psi|^2\vec{\nabla}S] = {m\Omega\over\hbar}
(x\partial_y -y\partial_x) |\psi|^2.
\end{equation}
$S_1$ and $S_2$ correspond to the same positions of the vortex
cores, so that their difference $S_{12}$ is a single-valued function
of the position, solving
\begin{equation}
\mbox{div}[|\psi|^2\vec{\nabla}S_{12}] = 0.
\end{equation}
To show that in the case of a trapped condensate $S_{12}$
is a constant we consider the following  integral
\begin{equation}
I = \int\!\!\int_{{\cal A}} \mbox{div}[|\psi|^2 S_{12} \vec{\nabla}S_{12}]
\end{equation}
where the integration runs over the area $\cal A$ of the condensate.

First, we transform $I$ using Gauss's formula into an integral over
the border $\bar{\cal A}$ of the condensate:
\begin{equation}
I =  \int_{\bar{\cal A}} |\psi|^2 S_{12} \vec{\nabla}S_{12}\cdot\vec{n} =0
\end{equation}
which vanishes as $|\psi|^2=0$ on the border of the condensate.

Second, we expand the integrand of $I$ as
\begin{equation}
\mbox{div}[|\psi|^2 S_{12} \vec{\nabla}S_{12}] = S_{12} \mbox{div}[|\psi|^2
\vec{\nabla}S_{12}] + |\psi|^2 (\vec{\nabla}S_{12} )^2.
\end{equation}
The first term in the right hand side vanishes in virtue of the continuity
equation. Therefore
\begin{equation}
0=I=\int\!\!\int_{{\cal A}} |\psi|^2 (\vec{\nabla}S_{12} )^2.
\end{equation}
As the integrand is positive this implies $\vec{\nabla}S_{12}=\vec{0}$,
that is $S_{12} = $constant.

\begin{figure}[htb]
\caption{Isocontours of the density $|\psi|^2$ for a
1-vortex (a) and a 4-vortex (b) configuration obtained numerically
in a non-axisymmetric trap ($\epsilon=0.3$) 
with a rotation frequency $\Omega=0.2\omega$, and $\mu\simeq 
40\hbar\omega$ (see text for the definition of $\epsilon,\omega$);
the unit of length for $x$ and $y$ is $(\hbar/m\omega)^{1/2}$.
\label{fig:aniso}}
\end{figure}

\begin{figure}[htb]
\caption{\label{fig:eeps} Numerical results for the energy of 0-vortex,
1-vortex and 4-vortex configurations (from top to bottom), as function of
the frequency ratio $\omega_x/\omega_y$, for a fixed
product of the frequencies $\omega^2=\omega_x\omega_y$. The rotation
frequency of the trap is $\Omega=0.2\omega$ and the energies
are measured in units of $\hbar\omega$ from the 0-vortex
energy $E_{iso}$ in the axi-symmetric trap; the chemical
potential is $\mu \simeq 40 \hbar\omega$.
The circles correspond to numerical results; 
the solid line is an analytical
prediction for the 0-vortex case 
as obtained in section \protect\ref{subsec:phase}.
}
\end{figure}

\begin{figure}[htp]
\caption{Isocontours of the total density $\rho_{1+2}$
for two ballistically expanded
condensates. The condensate 2 has two vortices; it was prepared
in an axi-symmetric trap with $\Omega=0.2\omega$. The condensate 1
has no vortex and is slightly displaced along the axis $x$ connecting
the two vortex cores at time $t$.\label{fig:interf}}
\end{figure}

\begin{figure}[htb]
\caption{\label{fig:self} 
Self-energy of a vortex in an axisymmetric
trap as function of the distance $\alpha R$
of the core from the trap center, for $\mu_0\simeq 80\hbar\omega$. 
(a) $\Omega=0.03\omega$ and (b) $\Omega=0.045\omega$.
The solid lines are given by the analytical prediction $W(\vec{\alpha})$.
The stars are obtained numerically.
The critical frequencies
defined in the text are $\Omega_c=2\Omega_{stab}\simeq 0.06 \omega$. 
$E_{iso}$ is the energy of the 0-vortex solution and the unit of energy
is $\hbar\omega$.}
\end{figure}

\begin{figure}
\caption{\label{fig:inter} Interaction energy between a vortex at the center of the trap
and a vortex at a distance $\alpha R$ from the center, as given
by the analytical formula Eq.(\protect\ref{INTER}) for $V$. The trap
is axisymmetric; the unit of energy is $(\hbar\omega)^2/\mu_0$.}
\end{figure}

\begin{figure}
\caption{\label{fig:deux} 
Energy of a system of two vortices symmetrically displaced
by $\pm\alpha R$ from the trap center, as function of the displacement
$\alpha$,
for $\Omega=0.1\omega$ and $\mu_0\simeq 80 \hbar\omega$. Solid line:
analytical result. Stars: numerics.  The trap
is axisymmetric; $E_{iso}$ is the energy of the 0-vortex solution 
and the unit of energy is $\hbar\omega$.
}
\end{figure}

\begin{figure}[htb]
\caption{\label{fig:diag}  
Local minima of energy
with $n$ vortices in an axisymmetric trap as function of the rotation
frequency $\Omega$ of the trap in units of the trap
frequency $\omega$. The chemical potential $\mu_0$ is approximately
$40\hbar\omega$. Note that for a fixed $n$ we
kept only the local minimum with the lowest energy.
The circles are numerical results.
The reference of energy 
is $E_{iso}$, the energy per particle in the absence of vortex, and
the unit of energy is $\hbar\omega$.
The lines with increasing absolute value of the slope correspond
to $n=1,\ldots,4$ vortices respectively.
}
\end{figure}

\begin{figure}[htb]
\caption{\label{fig:crit} Critical frequencies $\Omega_{stab}(n)$ and $\Omega_c(n)$
(see text) obtained from the analytical
Ansatz as function of the number of vortices. The chemical potential
$\mu_0$ is approximately $40\hbar\omega$ and the unit for $\Omega$
is the trap frequency $\omega$.
}
\end{figure}

\begin{figure}[htb]
\caption{\label{fig:n=6} In an axi-symmetric trap with $\Omega
=0.3\omega$ and $\mu_0\simeq 40\hbar\omega$, 
different configurations of $6$ vortices corresponding to a local
minimum of energy: (a) 6 vortices on a circle, with an energy
per particle $E=E_{iso}-0.5910\hbar\omega$. (b) one vortex at the center
and 5 vortices on a circle, with $E=E_{iso}-0.5890\hbar\omega$. 
$E_{iso}$ is the energy per particle in the absence of vortex.
The unit of length is $(\hbar/m\omega)^{1/2}$.
}
\end{figure}

\begin{figure}[htb]
\caption{\label{fig:n=18} In an axi-symmetric trap with $\Omega
=0.5\omega$ and $\mu\simeq 40\hbar\omega$,
different configurations of $18$ vortices corresponding to a local
minimum of energy: 
(a) with one vortex at the center
$E=E_{iso}-2.3988\hbar\omega$, this is the global minimum of energy.
(b) without vortex core at the center; the energy
per particle is slightly higher, $E=E_{iso}-2.3954\hbar\omega$. 
$E_{iso}$ is the energy per particle in the absence of vortex.
The unit of length is $(\hbar/m\omega)^{1/2}$.
}
\end{figure}

\begin{figure}[htb]
\caption{\label{fig:3d} A local minimum energy solution in 3D with 5 vortices
obtained from the numerical evolution in complex time.
The trapping frequencies are in the ratio $\omega_x=\omega_y=4\omega_z$.
The chemical potential is $53.7 \hbar\omega_{x,y}$. The trap is rotated
at a frequency $\Omega=0.25\omega_{x,y}$. (a) Isocontours for a cut
of the density in the plane $z=0$, showing the 5 vortex cores. (b) Isocontours
for a cut of the density in the $x-z$ plane, showing the dependence 
with $z$ of the vortex cores. The unit of length is
$(\hbar/m\omega_{x,y})^{1/2}$.}
\end{figure}


\begin{thebibliography}{99}
\bibitem{COMMENT} The numerical results for the energy curves
are obtained as follows. We start with $\psi$ given by the Ansatz for
one or several vortices 
displaced from the trap center and evolve it in complex time.
We observe that the positions of the vortices evolve very slowly
whereas the other variables such as the phase relax rapidly to a
quasi steady-state. This allows to define an energy
as function of the positions of the cores.
\end{thebibliography}
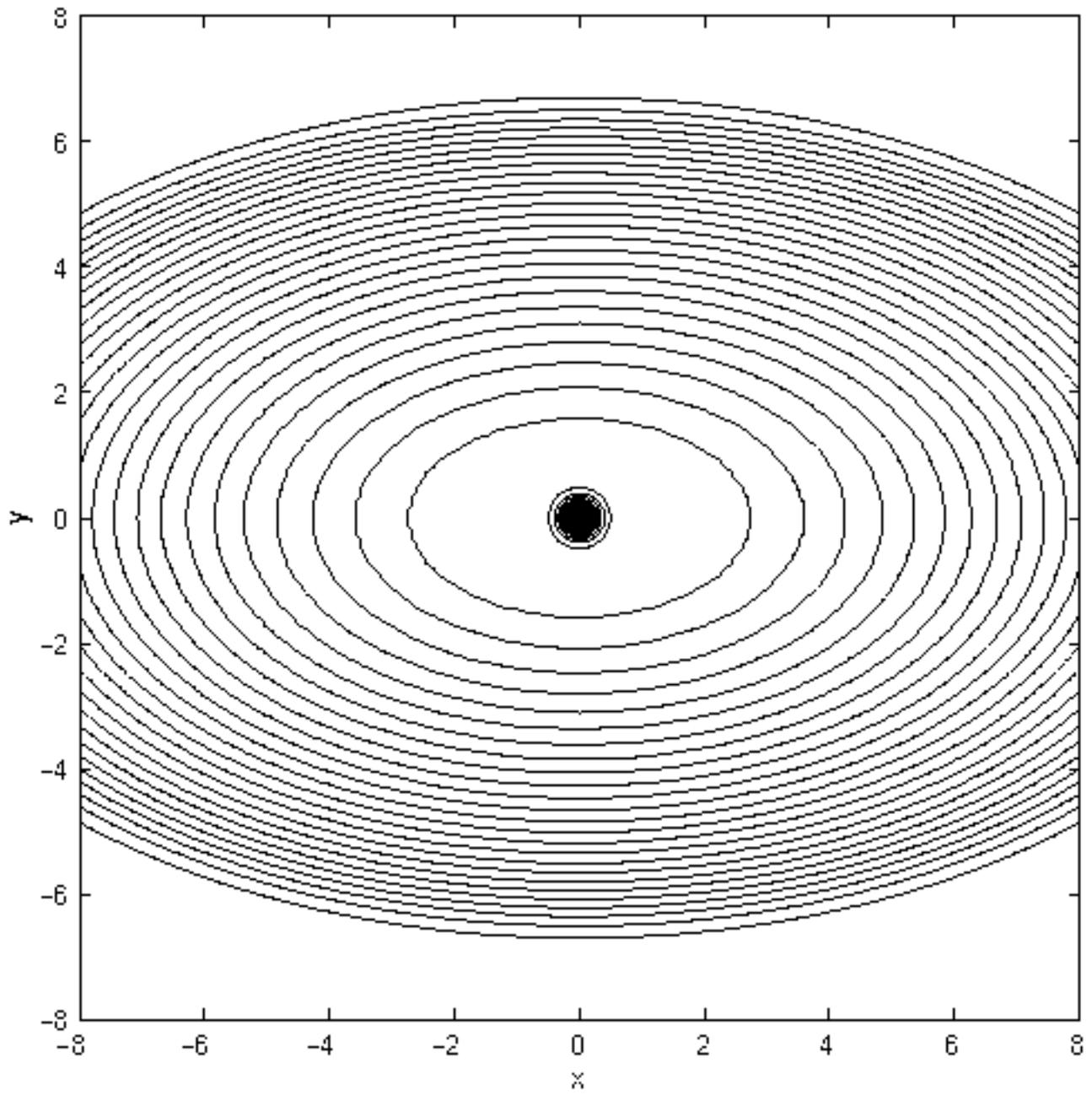
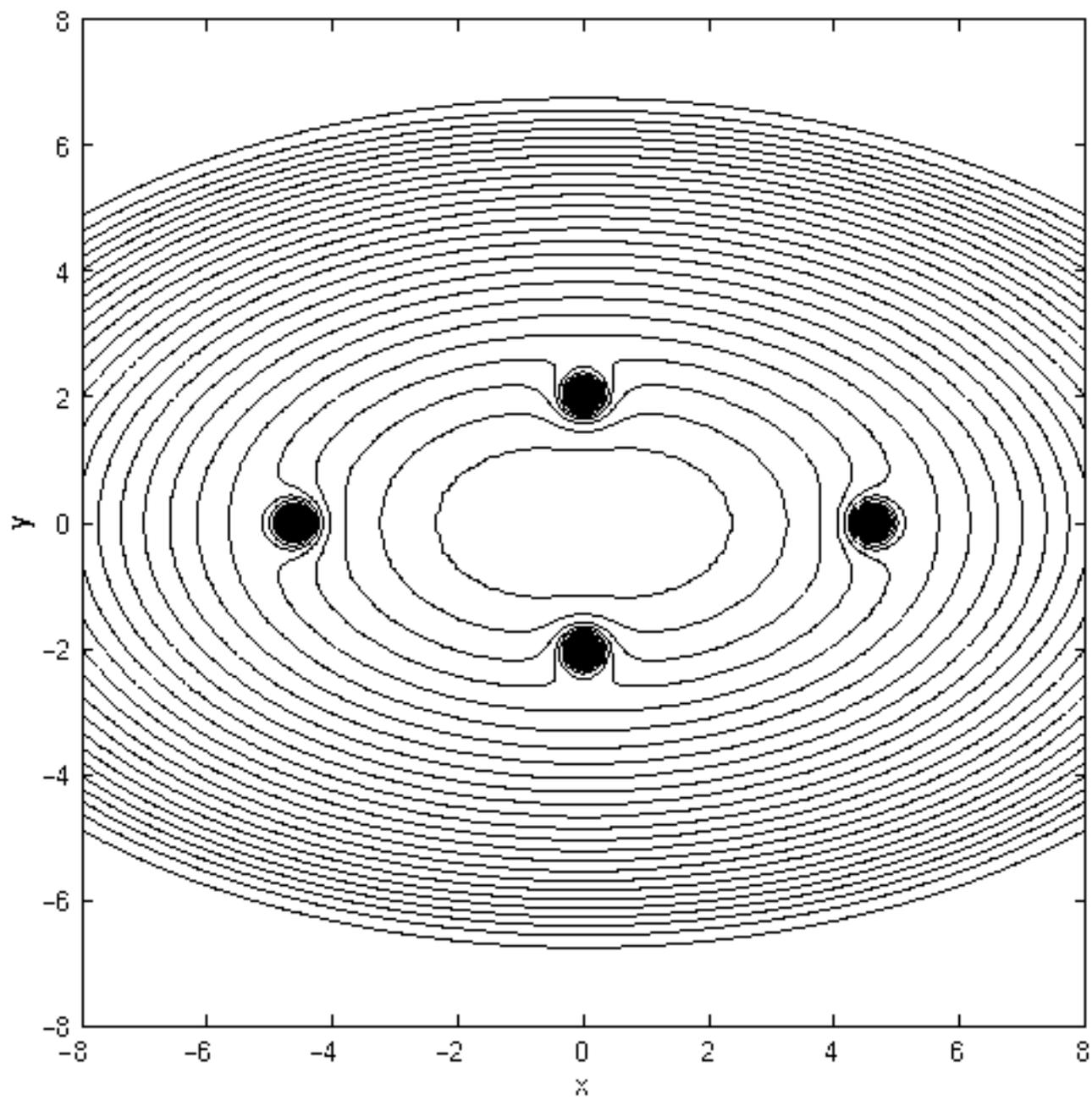
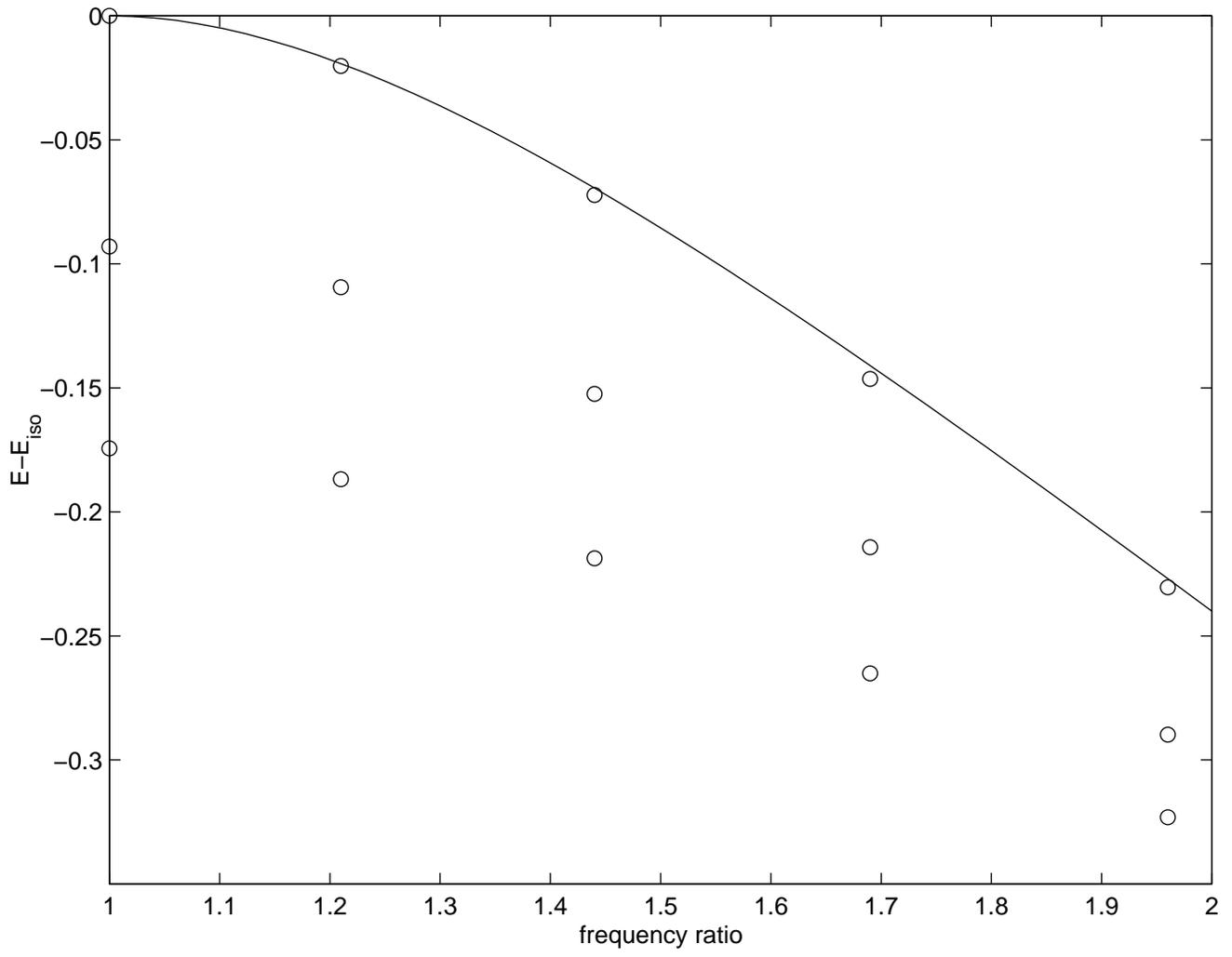
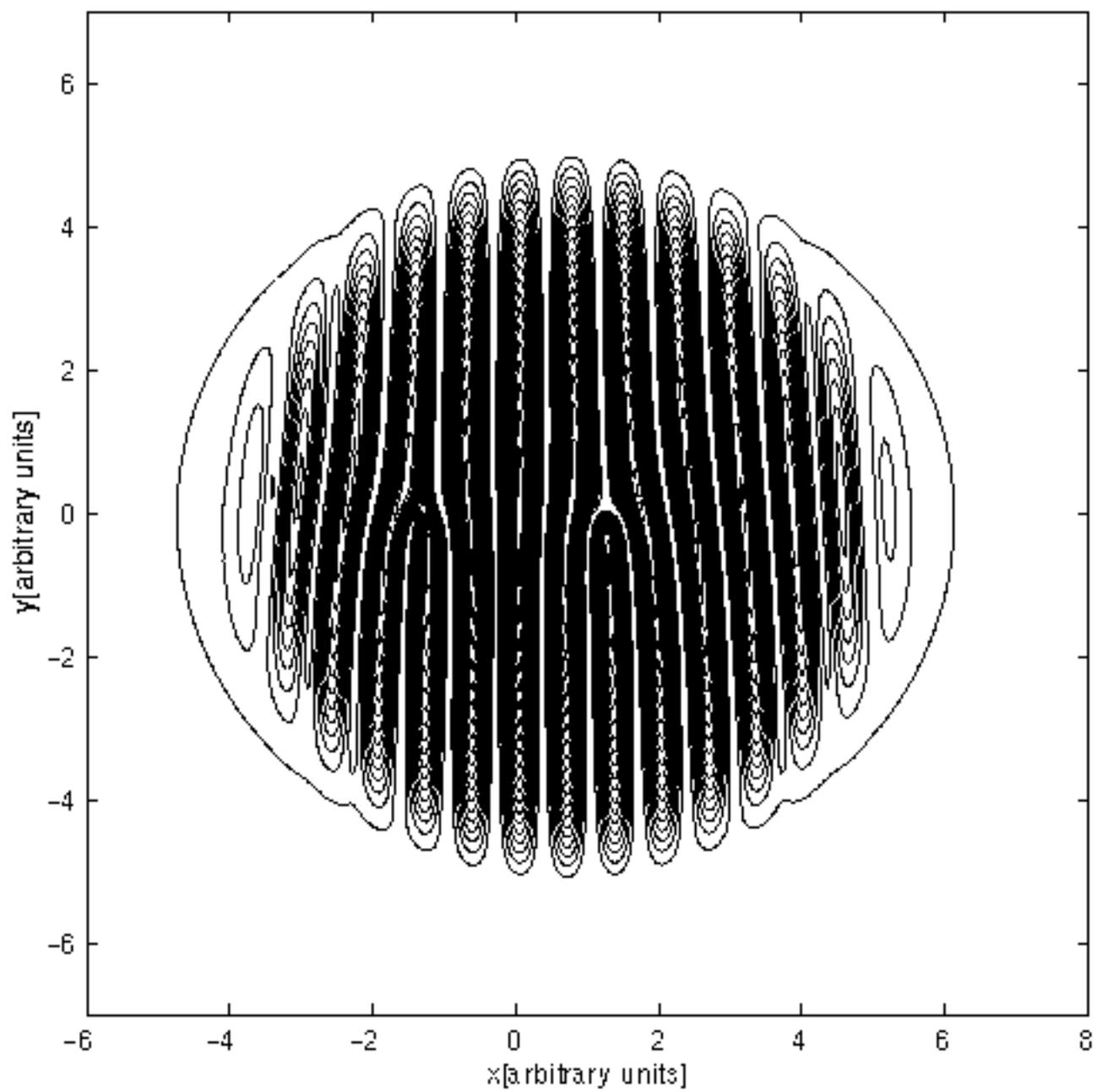
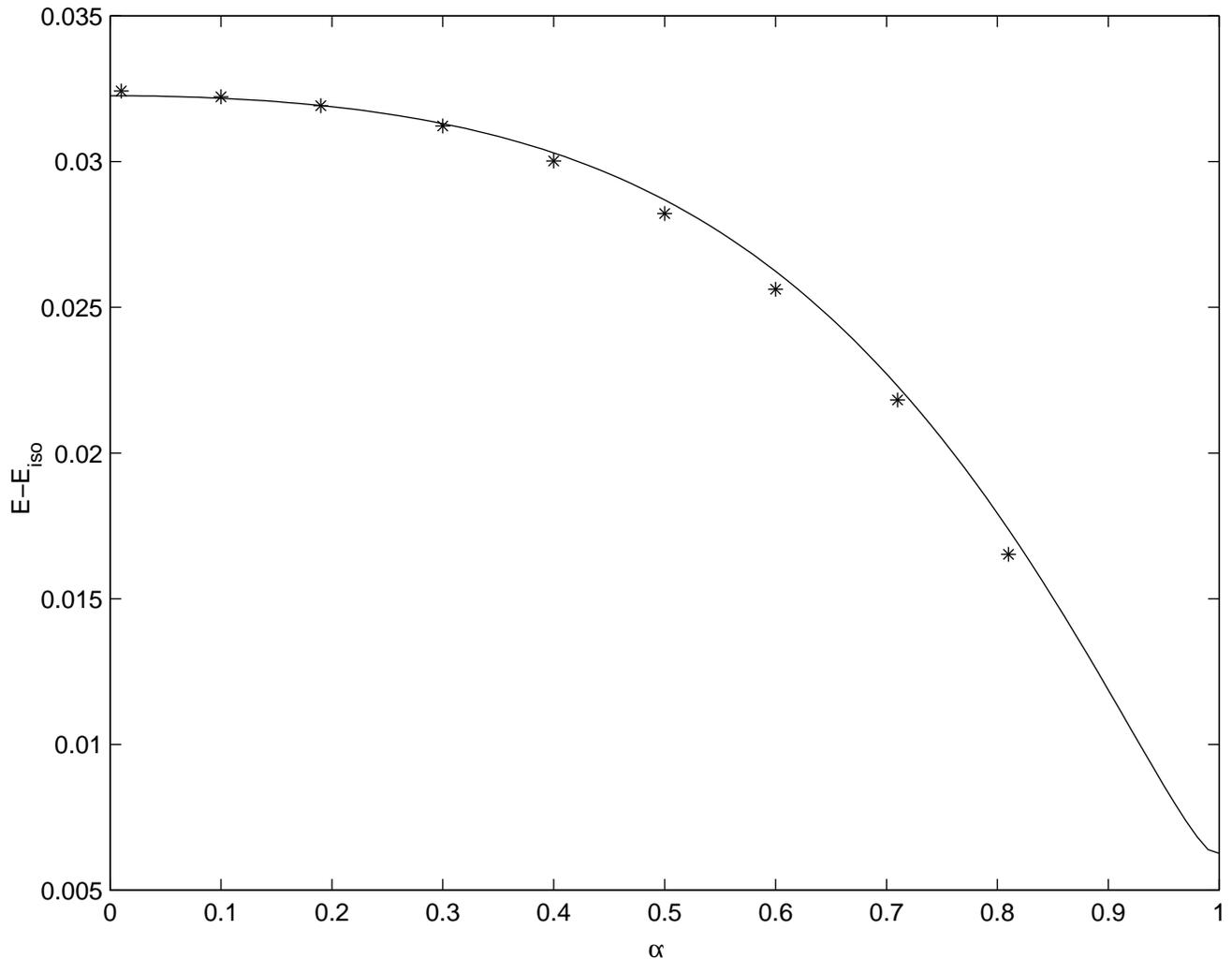
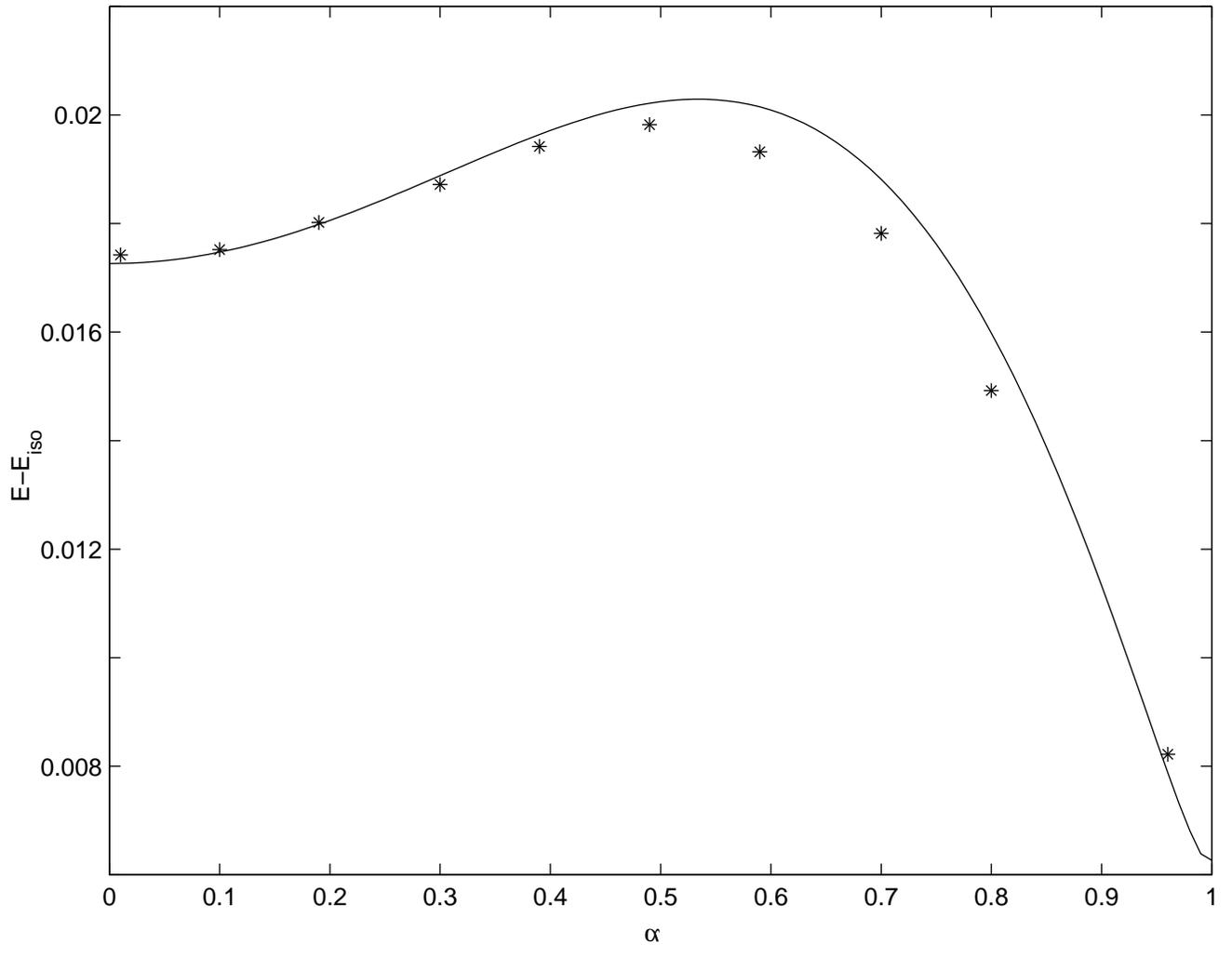
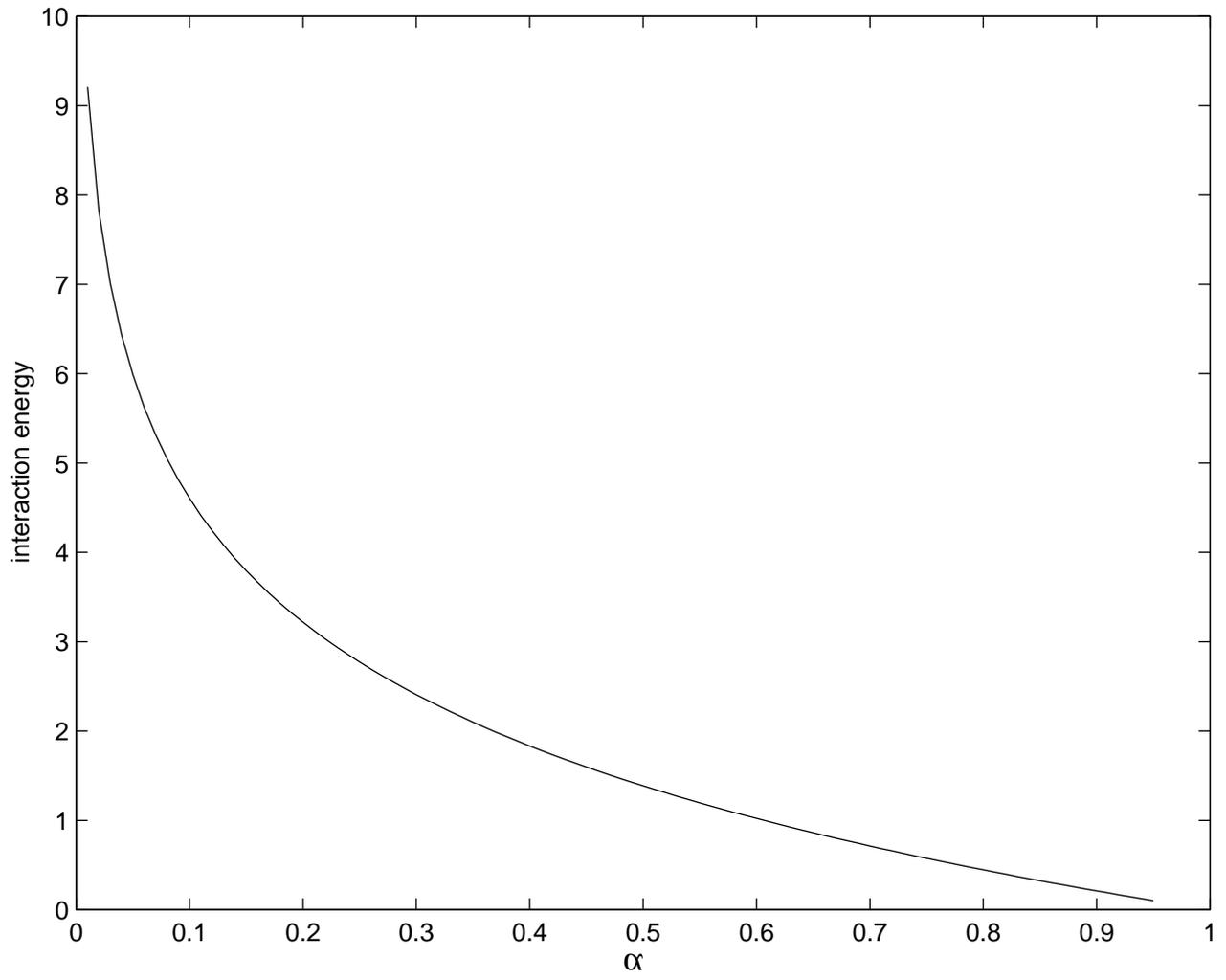
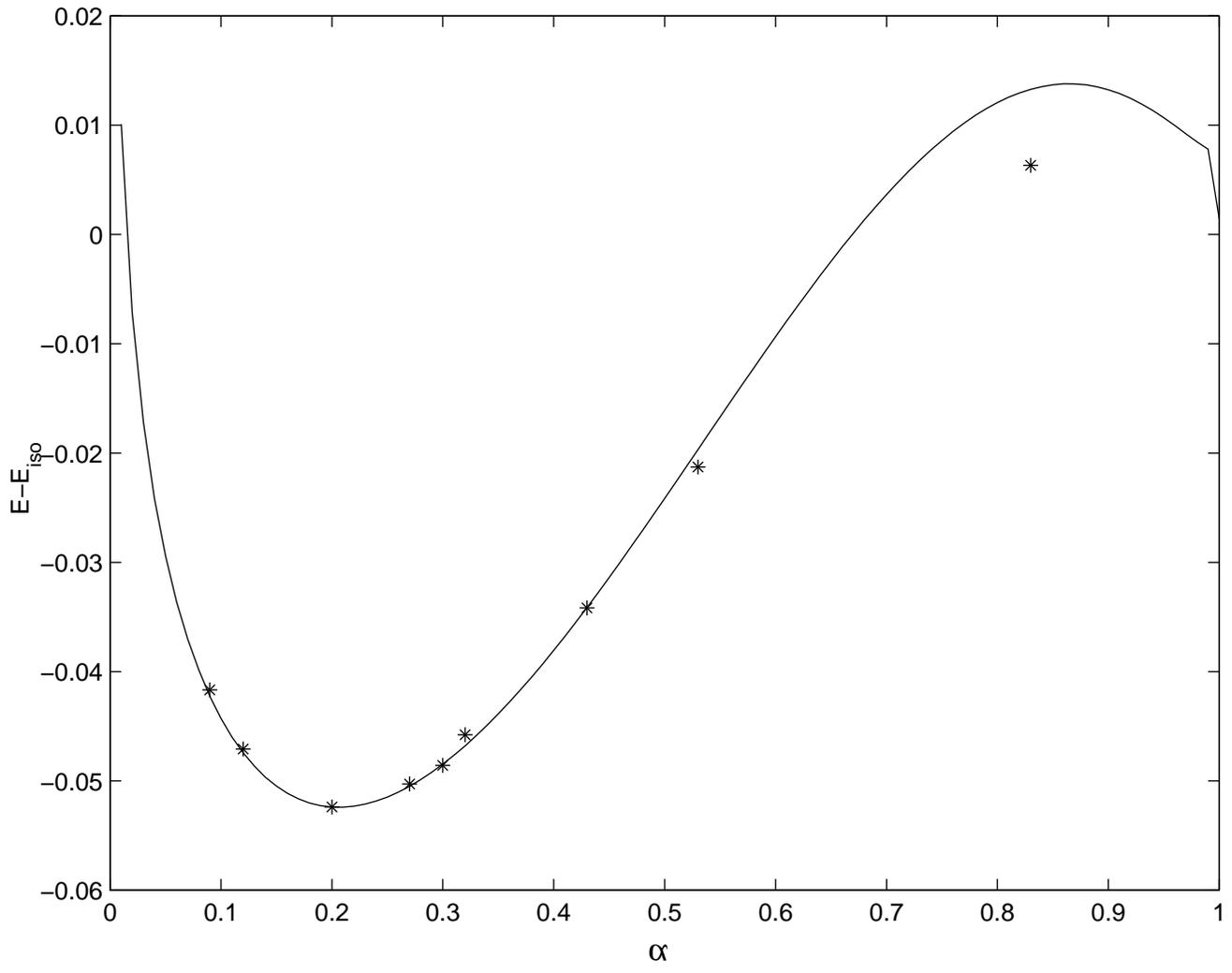
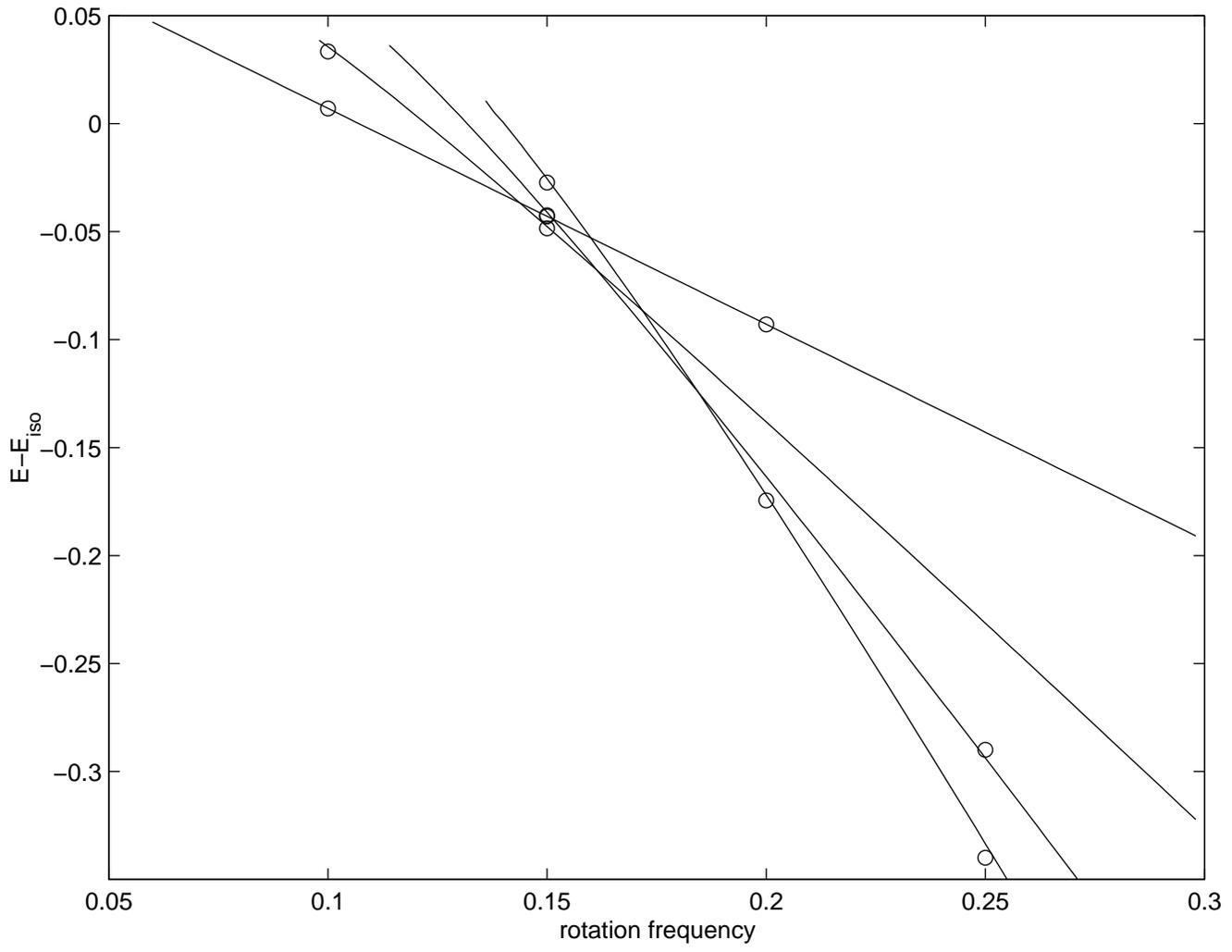
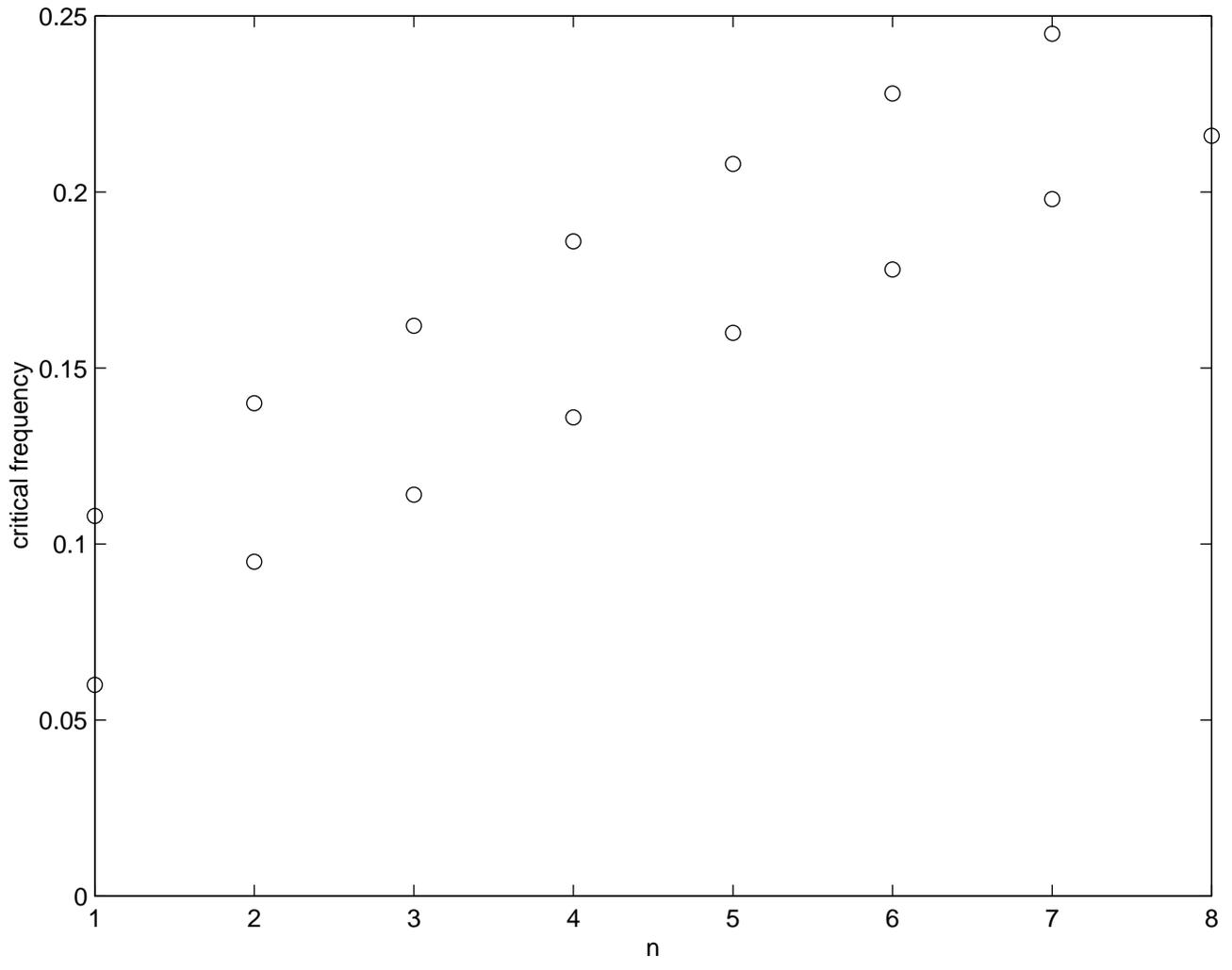
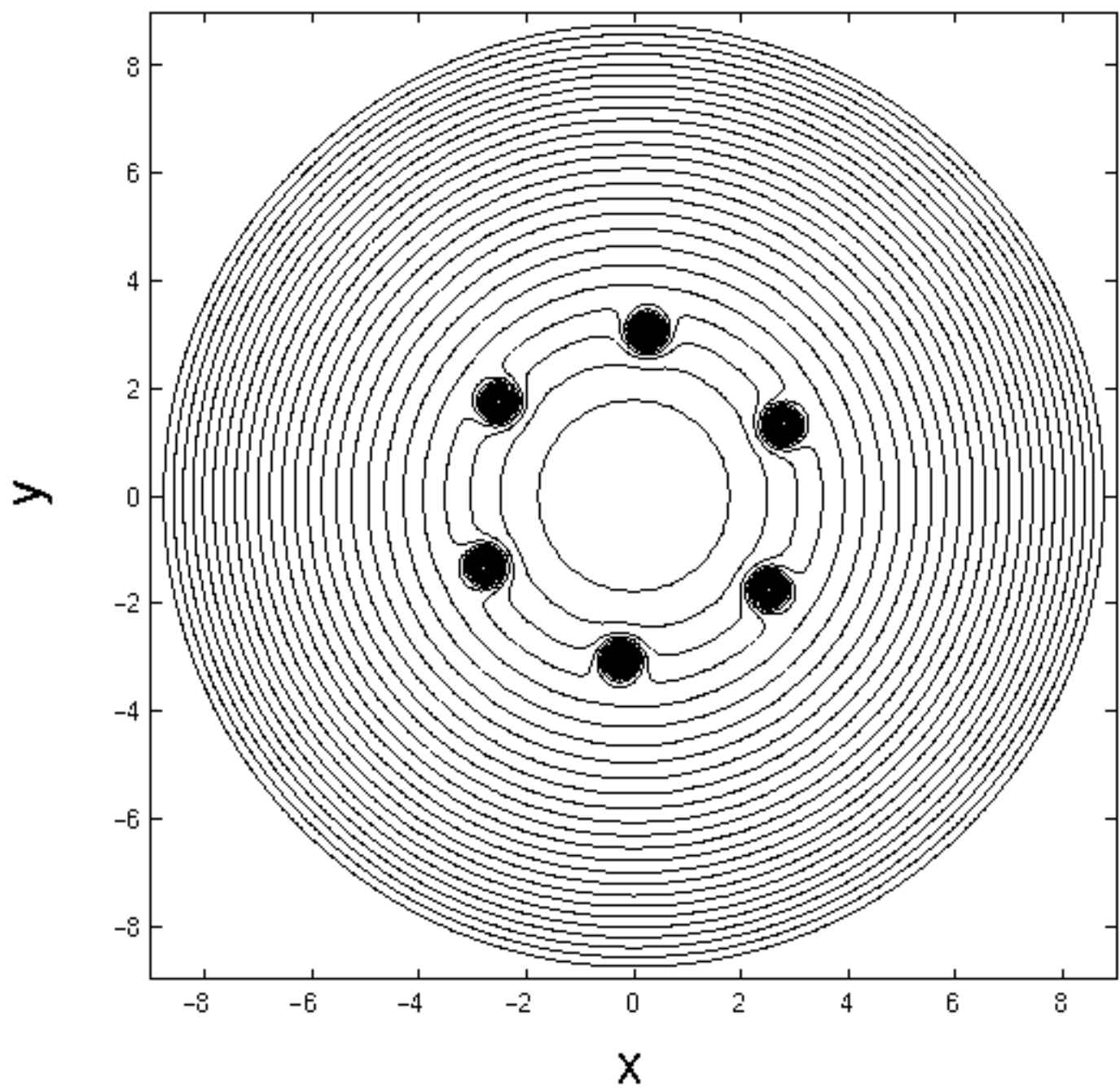
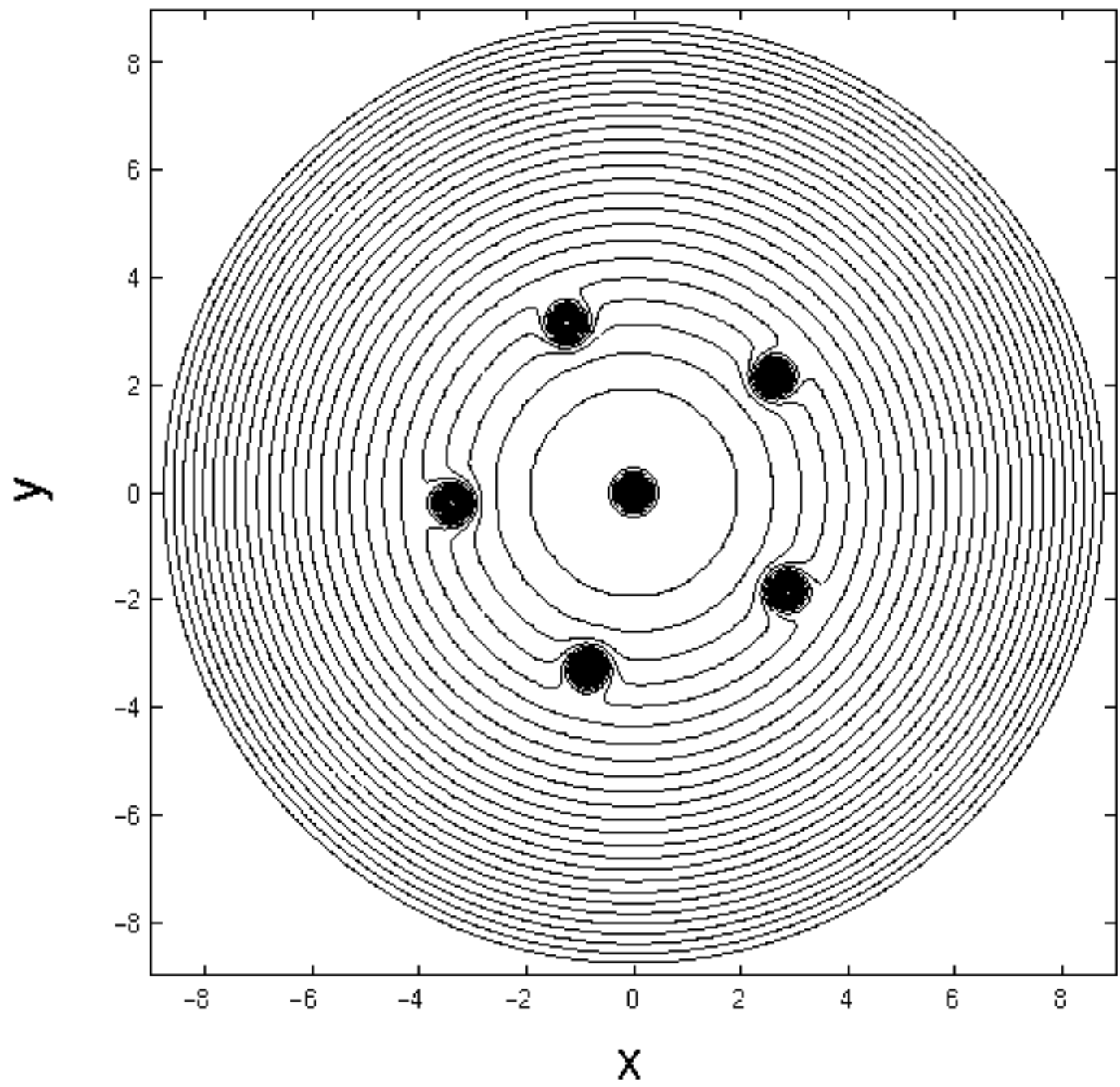
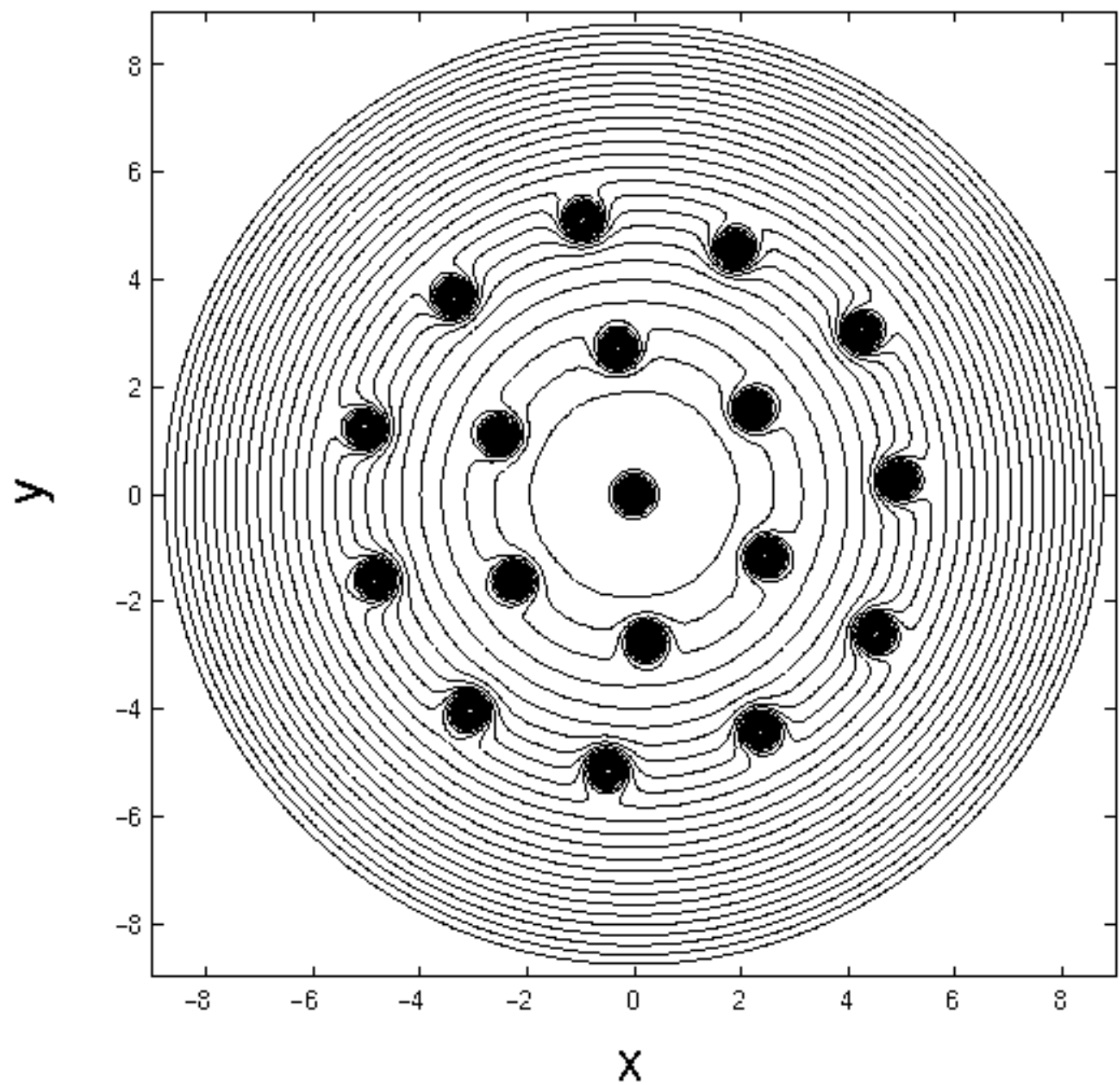
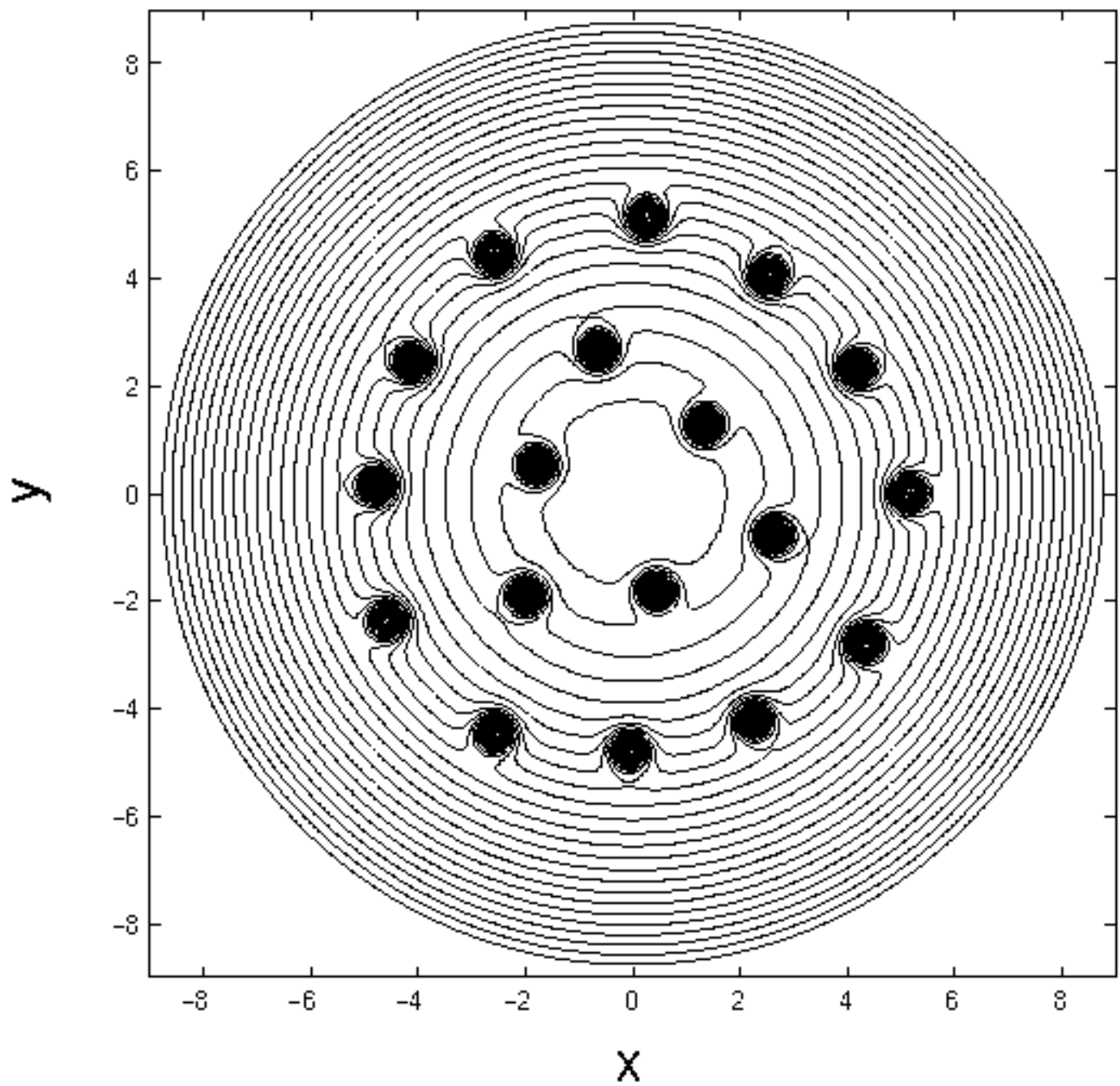
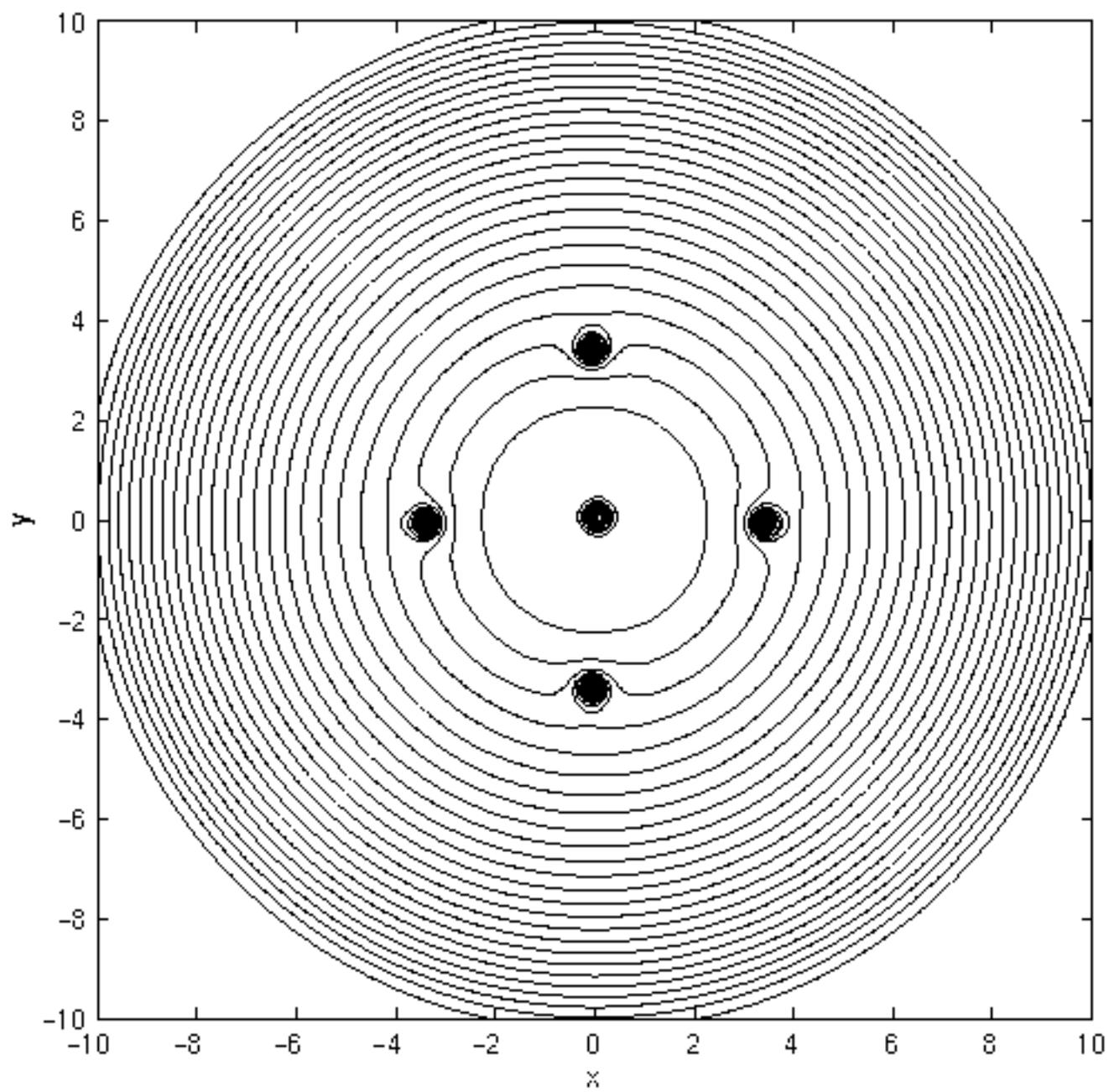
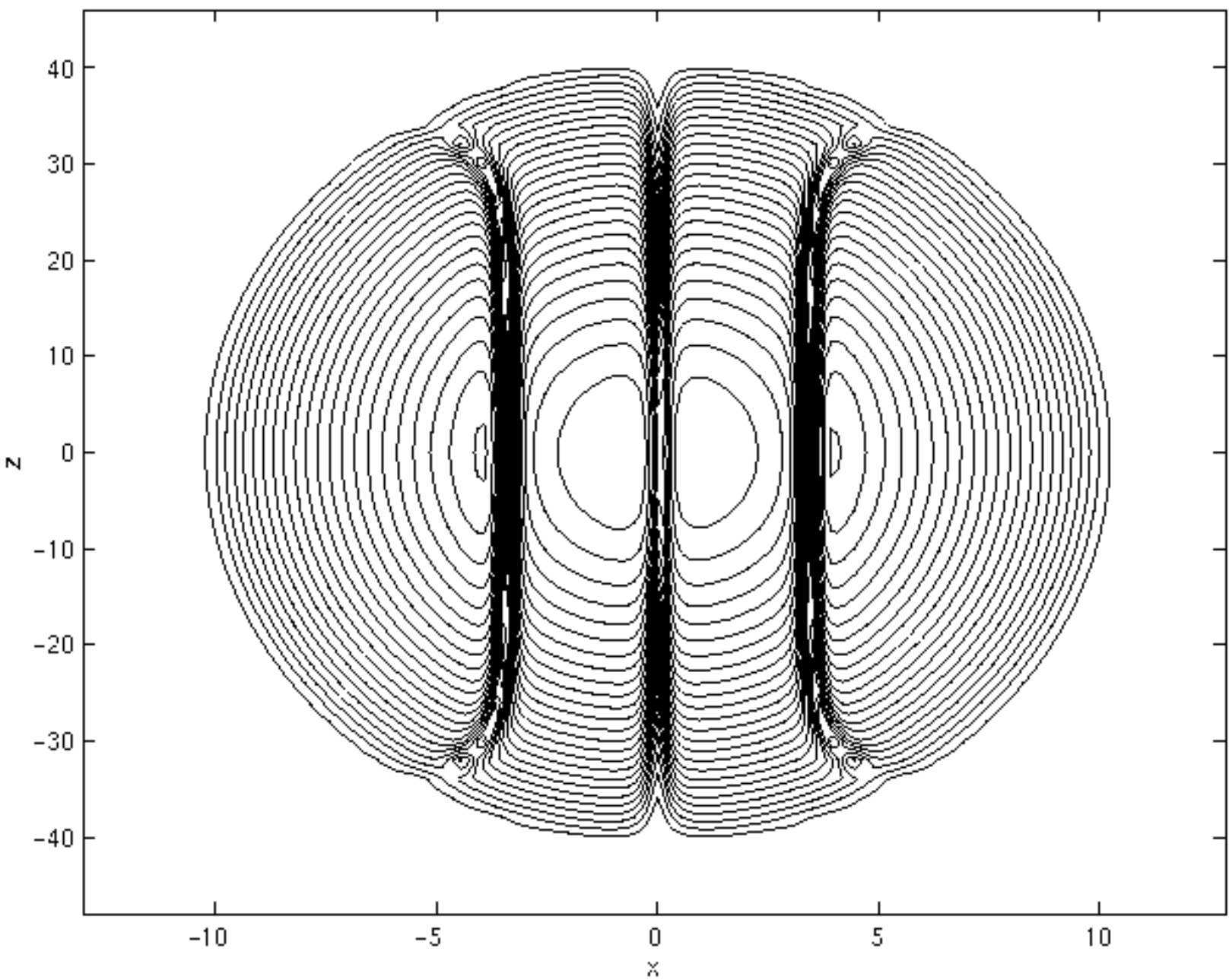
\end{document}